\documentclass{achemso}
\setkeys{acs}{email=false}
\usepackage[english]{babel}
\usepackage[T1]{fontenc}

\usepackage{csquotes}
\usepackage[fleqn]{amsmath}
\usepackage{amsmath}
\usepackage{setspace}
\usepackage{pdfpages}
\newtheorem{theorem}{Theorem}
\title{}

\begin{document}
\singlespacing
\vspace{-1.5in}
\begin{center}
	\begin{Large}
\textbf{Maximal predictability approach for identifying the right descriptors for electrocatalytic reactions}\\
	\end{Large}
\vspace{0.1in}
\begin{large}
\textbf{Dilip Krishnamurthy$^{a,\dagger}$, Vaidish Sumaria$^{b,\dagger}$, Venkatasubramanian Viswanathan$^{a,b,*}$}\\
\end{large}
\end{center}
\vspace{0.1in}
$^a$ Department of Mechanical Engineering, Carnegie Mellon University, Pittsburgh, Pennsylvania, 15213, USA\\
$^b$ Department of Chemical Engineering, Carnegie Mellon University, Pittsburgh, Pennsylvania, 15213, USA\\
$^\dagger$ Equal contribution\\
$^*$ Corresponding author, Email: venkvis@cmu.edu\\

\begin{abstract} Density Functional Theory (DFT) calculations are being routinely used to identify new material candidates that approach activity near fundamental limits imposed by thermodynamics or scaling relations.  DFT calculations have finite uncertainty and this raises an issue related to the ability to delineate materials that possess high activity.  With the development of error estimation capabilities in DFT, there is an urgent need to propagate uncertainty through activity prediction models.  In this work, we demonstrate a rigorous approach to propagate uncertainty within thermodynamic activity models.  This maps the calculated activity into a probability distribution, and can be used to calculate the expectation value of the distribution, termed as the expected activity.  We prove that the ability to distinguish materials increases with reducing uncertainty.  We define a quantity, prediction efficiency, which provides a precise measure of the ability to distinguish the activity of materials for a reaction scheme over an activity range.  We demonstrate the framework for 4 important electrochemical reactions, hydrogen evolution, chlorine evolution, oxygen reduction and oxygen evolution.  We argue that future studies should utilize the expected activity and prediction efficiency to improve the likelihood of identifying material candidates that can possess high activity.
\end{abstract}

Density functional theory (DFT) simulations are now being routinely used to guide material discovery in heterogenous electrocatalysis.\cite{norskov2009towards}  Electrocatalysis has seen numerous success stories for theory-guided material design through the use of descriptor-based searchers in hydrogen evolution\cite{greeley2006computational, greeley2006hydrogen, norskov2005trends}, oxygen reduction\cite{greeley2009combinatorial,viswanathan2012universality}, hydrogen peroxide synthesis\cite{rankin2012trends,viswanathan2015selective,verdaguer2014trends} and oxygen evolution \cite{man2011universality, halck2014beyond}.   The approach of all of these studies typically involves identifying the descriptors, typically adsorption energies of certain reaction intermediates, through the use of scaling relations and the identification of key potential-determining steps from the full reaction mechanism.   A large number of possible catalysts are subsequently examined and those that exhibit optimal values for the descriptors are used as candidates for further investigation.

Given the importance of the descriptor, a key missing piece is a robust theoretical basis for the selection of descriptors.  Due to the existence of scaling relations, there are often multiple descriptor choices, for e.g., for the ORR, adsorption free energy of O* \cite{norskov2004origin} or OH* \cite{viswanathan2012universality, deshpande2016quantifying}. In parallel, an emergent frontier in DFT is the incorporation of uncertainty associated with predictions.  The development of Bayesian Error Estimation Functional (BEEF) has brought in error-estimation capabilities to DFT simulations by generating an ensemble of functionals to map known uncertainties in the training datasets of the XC functionals.\cite{wellendorff2012density}

In this work, utilizing the uncertainty estimation capability, we argue that the choice of the descriptor must be the one that provides maximum distinguishability among the material predictions. Within this approach, the quantity of interest, such as the reaction rate or the limiting potential, becomes a probabilistic quantity and we show that the expectation value of the probability distribution exhibits certain unique properties.  In an earlier work, we defined a quantity termed the expected limiting potential, U$_{EL}$, and in this work, we prove that independent of the nature of the reaction scheme, the expected limiting potential, U$_{EL} \rightarrow $ U$_{L}$ as the uncertainty appears zero. We also define a quantity termed as the prediction efficiency ($\eta_{pred}$), which provides a quantification of the ability to distinguish the catalytic activity of materials for the property of interest, e.g. reaction rate or limiting potential. A quantity of interest for screening approaches emerges from this formalism, termed as the prediction limit, which represents the activity above which assertive predictions cannot be made using computational approaches alone due to indistinguishability. We demonstrate the framework for several important electrocatalytic reactions, (i) hydrogen evolution reaction, (ii) chlorine evolution reaction, (iii) oxygen reduction reaction (2e$^-$ and 4e$^-$) and (iv) Oxygen evolution reaction.  We show that the optimal descriptors for 4e$^-$ and 2e$^-$ ORR are $G_{OH^*}$ and $G_{OOH^*}$, respectively.  For oxygen evolution reaction, the optimal descriptor is $\Delta G_2 = \Delta G_{O^*} - \Delta G_{OH^*}$.  In general, we find that prediction efficiency for 2e$^-$ electrochemical reactions is greater than that for 4e$^-$ reactions.

\section{Results and Discussion}

\subsection{Probabilistic Formalism for Activity Prediction} \label{Theorems}
Computational screening for electrocatalysts involves identifying candidates that exhibit certain optimal values for the specific descriptor. In the proposed approach, we consider the descriptor as a probabilistic variable and an important question that arises is how this probability distribution affects the activity, which now is a probabilistic quantity.  Based on insights from probability and statistics, we observe that the expected value from the probability distribution of activity possesses important properties. 

Let us consider the descriptor to be a normal distribution, $X \sim \mathcal{N}(\mu,\sigma^2)$, with the standard Gaussian probability density function (PDF), $p_x(x\ |\ \mu,\sigma^2)$, and we will revisit this assumption in the next section.  The activity, now a probabilistic quantity, is given by $A=f(X)$. The associated PDF, $\hat{p}_a(a)$ can be written as
$\hat{p}_a(a)=\int_{-\infty}^{+\infty} p_x(x)\delta(f(x)-a)~dx$. 
The normalized PDF, ${p}_a(a)$, can be obtained subsequently, and the expectation value of the activity can be obtained as the probability density weighted average.
$E[A] = \int_{a_{min}}^{a_{max}} a\ p_a(a)~da$ 
In order to build intuition about the properties of the expected activity, we prove the following:

\begin{theorem}
	As $\sigma \rightarrow 0$ for the descriptor PDF, the expected value of activity, $E[A] \rightarrow f(\mu)$. 
\end{theorem}
    Given $\sigma \rightarrow 0$, this implies that $p(x) = \delta(x-\mu)$.
	
	$\implies p(a) = \int_{-\infty}^{+\infty} \delta(x-\mu)\delta(f(x)-a)dx = \delta(f(\mu)-a)$ 
	
	$\implies E[A] = \int_{a_{min}}^{a_{max}} a~\delta(a-f(\mu)) da = f(\mu)$ 
	
	$\implies \underset{\sigma \to 0}{\lim} E[A] = f(\mu)$

\begin{theorem}
	\label{th:concave}
	When the functional relationship between the descriptor and the activity, $A=f(X)$ is concave, $f(\mu) > E[A]$
\end{theorem}
	
	\noindent A concave function, $f(x)$ obeys $f(\sum \alpha_ix_i) > \sum(\alpha_if(x_i)) $ for $\sum \alpha_i = 1$. 
	
	\noindent Choosing $\alpha_i$ such that $\alpha_i = p_x(x)$, 
	$\implies f(\sum p_ix_i)> \sum(p_if(x_i)))$ 
	
	\noindent Since, by definition, $\sum p_ix_i = \mu$, and $\sum(p_if(x_i)))=E[A]$
	$\implies f(\mu) > E[A]$

\begin{theorem}
	At the maximal functional value, $a_{max}$, $E[A] < a_{max}~\forall~f(X)$.  
\end{theorem}
	
		\noindent By definition, $f(x) < a_{max} ~\forall~x$.  Multiplying by the positive quantity, $p_x(x_i)$ on both sides, we get
	
	$\implies p_x(x_i)f(x_i) < p_x(x_i)a_{max}$.  Now summing this up for all $i$, we get
	
	$\implies \sum p_x(x_i)f(x_i) < \sum p_x(x_i) a_{max}$.  Since $a_{max}$ does not depend on $i$, we get 
	
	$\implies E[A] < a_{max}$\\

In order to utilize these theorems, the descriptor PDF needs to be determined, which is enabled by the recent development of the Bayesian error estimation framework to the XC functionals\cite{wellendorff2012density}.  We demonstrate the use of error-estimation capabilities within the BEEF-vdW XC functional for obtaining the PDFs of descriptors through an ensemble of functionals. This approach has been utilized to quantify uncertainty associated with reaction energies,\cite{medford2014assessing, christensen2016functional} mechanical properties,\cite{PhysRevB94064105} and magnetic ground states.\cite{houchins2017quantifying} With the descriptor PDF determined, a key question that arises is how does the predictability with finite uncertainty compares to the case with no uncertainty, which we term as oracle (perfect) computation.  Distinguishability can be understood as the ability to delineate the activity difference of different materials.  Typically, in electrocatalysis, we are interested in identifying materials that possess a certain threshold activity, which in computational electrocatalysis is typically the threshold limiting potential, U$_{T}$.  This leads to a finite interval of the descriptor values of interest.  For this range of descriptor values, the activity with perfect computation (no uncertainty) in the descriptor space maps to an interval [U$_{T}$, $\max($U$_{L}$)], while that with finite uncertainty for the descriptor maps to [U$_{T}$, $\max($U$_{EL}$)], where $U_{EL}$ is the expectation value of the limiting potential.  An obvious approach to delineate materials is directly quantified by the length of this interval.  A mathematical precise definition of the ability to distinguish materials is the Lebesgue measure of the interval, which can be used readily for predicting more than one property of interest.  

Based on this, a quantity, which we term as the prediction efficiency, can be defined as the ratio of distinguishability with finite uncertainty to distinguishability with perfect computation, i.e. oracle computation.  Notationally, this is given as

$$\mathrm{\eta_{pred}(U_{T}) = \frac{\lambda([U_{T},\max(U_{EL})])}{\lambda([U_{T},\max(U_{L})])}}$$\\
\noindent where, $\lambda$ is the Lebesgue measure of the interval, which in one-dimension is its length.   We can build intuition on prediction efficiency based on the following properties.  With this definition, as $\sigma \rightarrow 0$ for the descriptor PDF, the prediction efficiency $\eta_{pred} \rightarrow 1$.  If $f$ is concave, based on Theorem \ref{th:concave}, the prediction efficiency, $\eta_{pred} < 1$ for $\sigma \neq 0$.  As we will show later, the prediction efficiency can be used to (i) quantify the efficiency of a particular descriptor for an electrocatalytic reaction scheme and (ii) quantitatively compare predictability between different electrochemical reactions.

\subsection{Hydrogen Evolution Reaction}\label{sec:HER}
\begin{figure}[H]
	\centering
	\includegraphics[width=\textwidth]{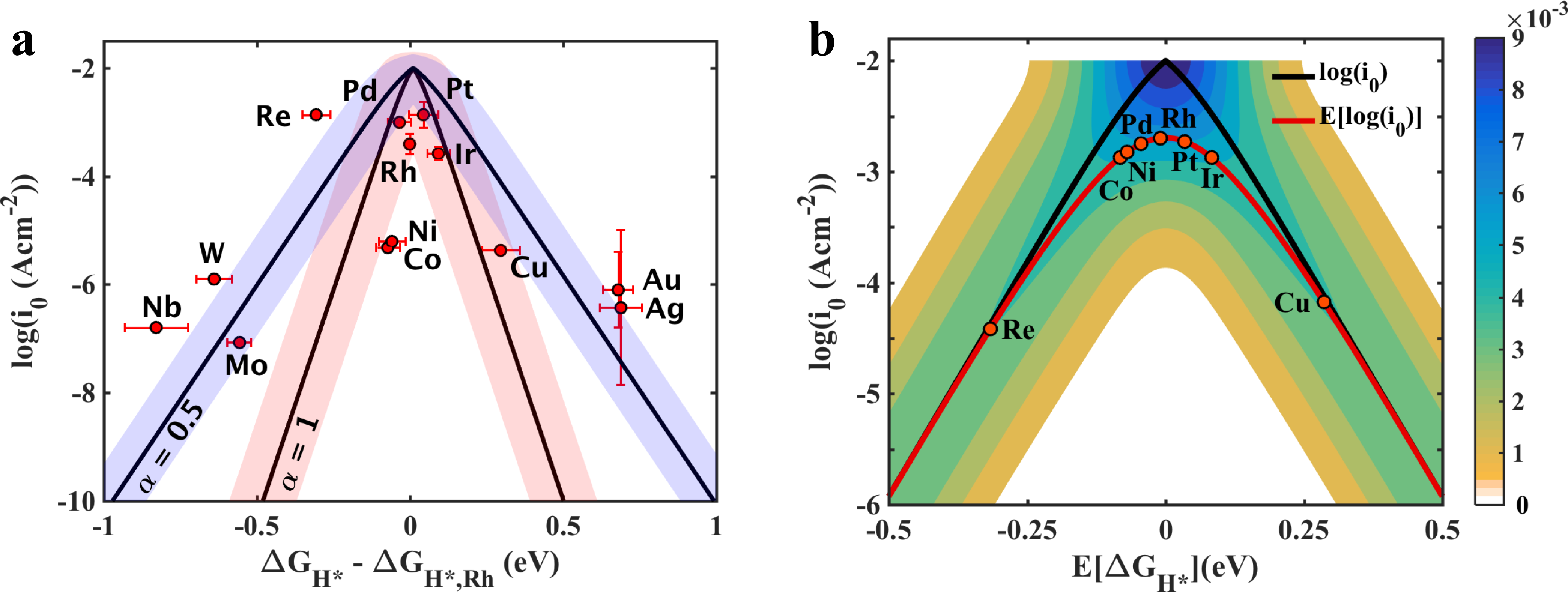}
	\caption{\textbf{Probabilistic activity prediction for the hydrogen evolution reaction based on transition metal (111) surfaces.} \textbf{a,} Exchange current density prediction with one standard deviation (shaded regions) from the kinetic model (black) based on the computed hydrogen binding energy and the experimentally measured activity (red dots) as compiled by  N{\o}rskov et al. \textbf{b,} Probability map of activity as a function of the expectation value of the hydrogen adsorption energy, computed using the outlined probabilistic uncertainty propagation framework outlined. The red line represents the expected activity, depicting reduced distinguishability of materials along the activity axis, which determines the ability of DFT to delineate the activity of material candidates. The area between the red and black lines represents the region of computational uncertainty, implying that finding candidates in this region (highly active candidates) requires higher-order computation and a synergistic experiment-computational effort.}
	\label{fig:HER}
\end{figure}

We begin with hydrogen evolution to illustrate our probabilistic approach owing to a unique well-established atomic-scale descriptor for its catalytic activity \cite{parsons1958rate}. The HER has gained renewed interest in solar water-splitting for hydrogen production \cite{lewis2006powering, mccrory2015benchmarking} requiring computational-screening approaches for identifying active catalysts. For prediction of HER activity before DFT-enabled computations, Parsons identified through a mechanistic understanding that the descriptor is the free energy of hydrogen adsorption. However, due to the then calorimetric limitations, for decades the descriptor was approximated to be the metal-hydride bond strength \cite{parsons1958rate}. With the development of DFT, it has been made possible to compute the hydrogen binding energy ($\Delta G_{H}$) and in agreement with experimental measurements, which has been used in numerous successful screening approaches \cite{greeley2006computational, greeley2006hydrogen, norskov2005trends}. In this work, we use this descriptor to demonstrate our probabilistic formalism for activity prediction by following the Volmer-Heyrovsky reaction mechanism \cite{bockris1959kinetics}. In this mechanism, the first step is the activation of protons as adsorbed hydrogen, and in a subsequent step undergoes a concerted proton-electron addition to evolve hydrogen. The descriptor, the hydrogen binding energy, and the activity, can be linked through a simplified kinetic model \cite{norskov2005trends} as $$i_0 = f(\Delta G_{H^*})=-ek_0(1+\exp(|\Delta G_{H^*}|/kT))^{-1}$$ where the pre-exponential factor is obtained by fitting to experiments. To demonstrate the probabilistic approach, we consider a range of transition metals that are known to be active hydrogen-evolution catalysts and the predicted exchange current density is shown in fig. \ref{fig:HER}(a). Within our uncertainty-propagation framework, as presented in an earlier work we approximate the descriptor uncertainty to be uniform and given by the standard deviation of the combined distribution of the descriptor, $\sigma_H$, from all the metals \cite{deshpande2016quantifying}. We treat the descriptor as a Gaussian probabilistic variable, $X \sim \mathcal{N}(\mu,\sigma_H^2)$, with an associated PDF, $p_x(x|\mu,\sigma_H^2)$. To propagate the descriptor uncertainty, we map this PDF through the kinetic model onto the exchange current density axis, and the activity PDF can be expressed as 
$$\hat{p}_{i_0} (i_0) = \int_{-\infty}^{+\infty} p_x(x) \ \delta(f(x) - i_0)dx$$ 
which is normalized to obtain $p_{i_0} (i_0)$, the activity PDF. Fig. \ref{fig:HER}(b) shows the PDF map of the activity as a function of $\mu$, the mean value of the descriptor PDF. From the activity PDF, the expectation value of activity is obtained using the normalized PDF $p \ (i_0)$ as $${i_0}_E = \int_{-\infty}^{{i_0}_{max}} i_0 \ p_{i_0}(i_0) \ \mathrm{d}i_0$$
The red curve in Fig. \ref{fig:HER}(b) represents the expected-activity curve and the distinguishability of activity between candidates near the top of the volcano is the lowest. This implies that through a purely computational approach, with the current DFT accuracy, the predicted activity of candidates like Pt, Pd and Rh are indistinguishable. Therefore, efforts with a quest to identify catalysts in acidic media better than platinum, the archetypical HER catalyst, must be cautious in choosing purely computationally-driven approaches.

\subsection{Chlorine Evolution Reaction}\label{sec:ClER}
\begin{figure}[H]
	\centering
	\includegraphics[width=\textwidth]{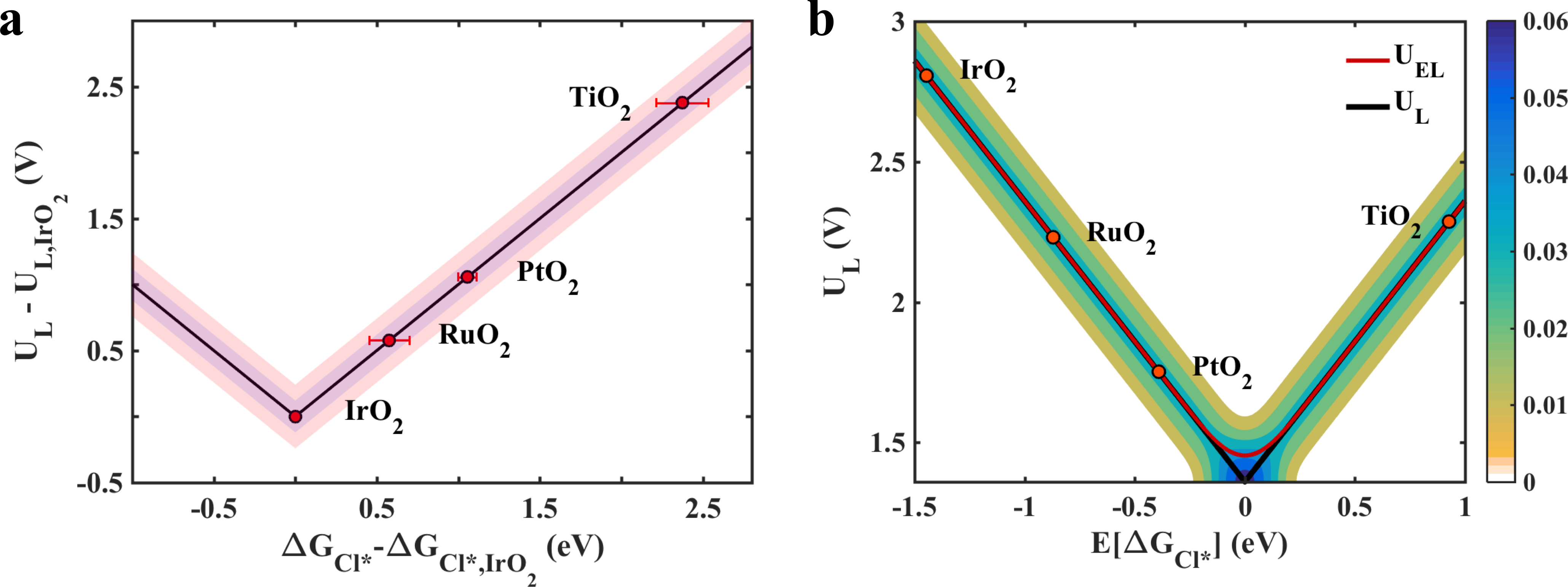}
	\caption{\textbf{The limiting potential for chlorine evolution predicted based on the probabilistic framework utilizing error estimation capabilities.} \textbf{a,} Limiting potential as a function of free energy of chlorine adsorption referenced to IrO$_2$ to minimize systematic errors in DFT. The solid line represents the predicted value from the volcano relationship whereas the colored regions represent uncertainty (1-$\sigma$ and 2-$\sigma$ regions). \textbf{b,} Probability distribution of the limiting potential plotted versus the expectation value of the free energy of chlorine adsorption. The red line represents the expected limiting potential, which is expected to better agree with the average activity from a very large number of experiments.}
	\label{fig:ClER}
\end{figure}

The chlorine evolution reaction (ClER), which is one of the largest technological application of electrochemistry, occurs as a 2-electron process through a few possible pathways with a well-studied descriptor for its catalytic activity. While the ClER is a hugely important reaction, the search for effective catalysts has largely been driven by empiricism. Through an empirical approach, the development of Dimensionally Stable Anodes by Beer\cite{beer1977method} forms a milestone for the chlor-alkali industry demonstrating a synergistic enhancement of stability and activity over a broad range of operating conditions\cite{trasatti1984electrocatalysis}. The mechanistic understanding of the ClER was largely driven by experimental work by Krishtalik et al.\cite{Krishtalik1981kinetics} However, oxygen evolution tends to occur as a parasitic reaction, especially at high current densities, since the equilibrium potentials for chlorine evolution and oxygen evolution are close and rutile oxides catalyze both reactions \cite{hansen2010electrochemical}. Undoubtedly, the possible competing pathways affect the specific activity, however, we adopt a simplistic mechanism to demonstrate our probabilistic framework by following the Volmer-Heryrovski mechanism. Since the mechanism involves a single intermediate, the chlorine adsorption energy forms the descriptor for activity and it is possible to attain the equilibrium potential. The limiting potential can be determined\cite{hansen2010electrochemical} as $U_L = f(\Delta G_{Cl^*})=1.36+|\Delta G_{Cl^*}|$ V . We consider rutile oxides that are reported to be active chlorine-evolution catalysts (fig. \ref{fig:ClER}(a)). We follow a similar approach to that demonstrated for HER, to propagate the descriptor uncertainty ($\sigma_{Cl}$) to the activity (Fig. \ref{fig:ClER}(b)), where the red curve represents the expected activity. The distinguishability of activity between candidates with descriptor values near the apex of the volcano is the lowest, as implied in theorem 2. The reduced distinguishability can be attributed solely to the descriptor uncertainty since there exists only a single intermediate (descriptor), which also results in a high prediction efficiency for ClER relative to reactions with multiple descriptor choices, as we demonstrate later (Fig. \ref{fig:compare}). We observe from the prediction efficiency curve that for overpotentials below ~0.9 V, the prediction efficiency is zero, implying that higher-order DFT methods are necessary for screening approaches to identify candidate catalysts with very low overpotentials.

\subsection{Oxygen Reduction Reaction}
The fundamental understanding of the ORR has largely been through the surface science approach to electrocatalysis\cite{markovic2002surface,mukerjee1995role,markovic1995oxygen,zhang2005controlling,zhou2009improving,kuzume2007oxygen,wakisaka2009identification,wakisaka2011structural,kondo2009active,stephens2011tuning}, which relies on surface analytical tools\cite{wakisaka2009identification,stephens2011tuning,damjanovic1967mechanism,fernandez2003scanning,hoster2004catalytic,stamenkovic2007improved} complemented by first-principles calculations\cite{norskov2004origin,anderson20022,viswanathan2012simulating,gohda2009influence,janik2009first,anderson2010solvation,jinnouchi2012first}. For computational screening, multiple choices of the descriptor for ORR activity have been used based on free energy scaling between intermediates \cite{greeley2009alloys, deshpande2016quantifying}. Greeley et al. showed that specific Pt-based binary alloys exceed the activity of Pt by using the free energy of adsorbed oxygen as the descriptor\cite{greeley2009alloys}. In contrast, Ifan et al. identified Pt$_5$La to have significantly higher activity than Pt by using the free energy of OH* as the descriptor \cite{stephens2012understanding}. A rational approach to choosing the right descriptor for the ORR is not present and in this work, we fill this gap with conclusions, similar to that shown in an earlier work\cite{deshpande2016quantifying}.  

\begin{figure}[H]
	\centering
	\includegraphics[width=\textwidth]{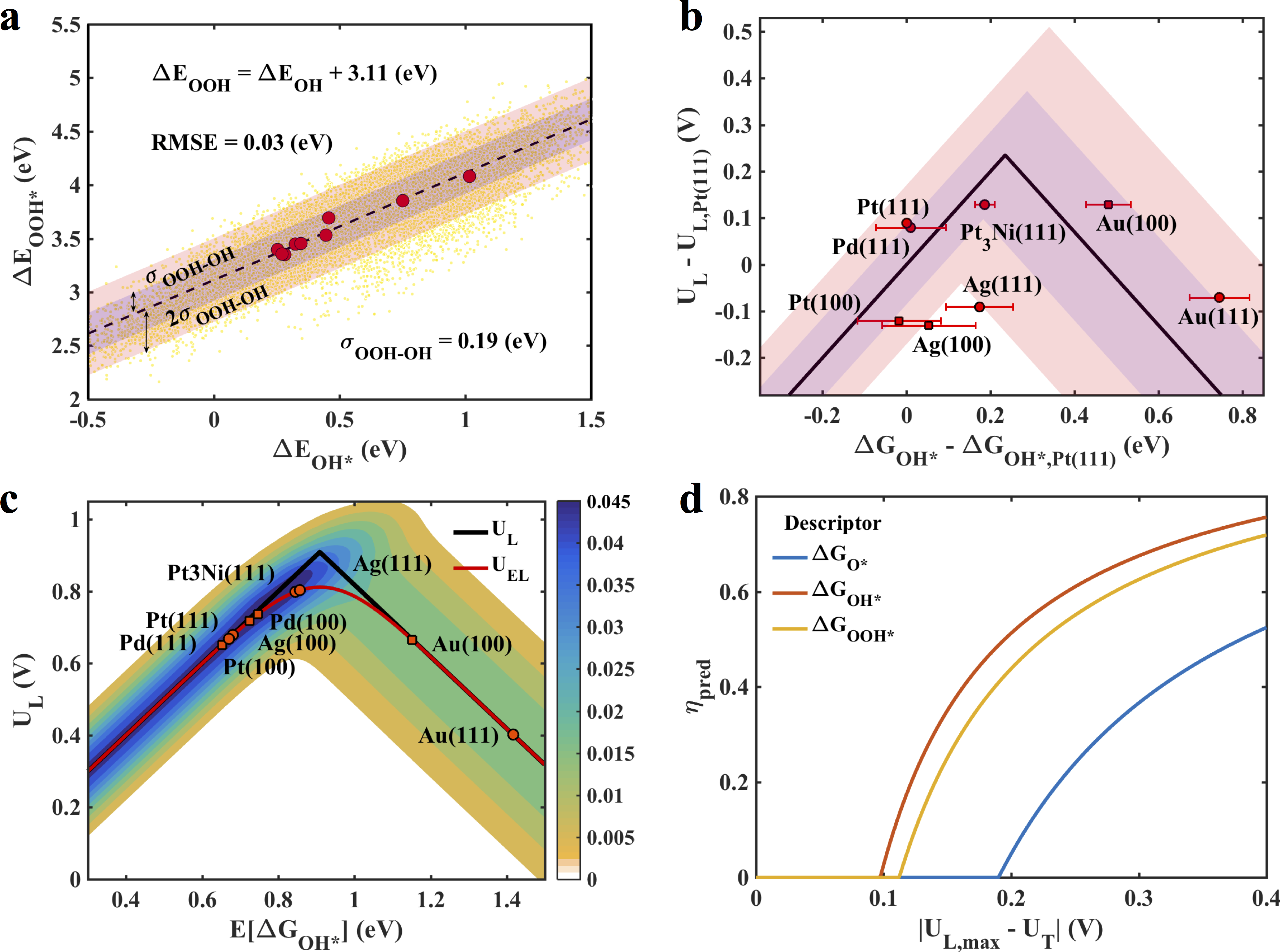}
	\caption{\textbf{Probabilistic ORR activity prediction using the identified maximal predictability descriptor. a,} Computed scaling relation between adsorption energies of the intermediates, OH$^*$ and OOH$^*$. The red dots represent DFT-calculated free energies and the yellow dots represent those computed from the family of functionals within BEEF-vdW, enabling error estimation. The two shaded regions represent one and two standard deviations in the scaling intercept. Similar scaling is found between the adsorption energies of other oxygen intermediates (shown in the Supporting Information). \textbf{b,} Activity volcano for 4e$^-$ ORR showing the experimentally measured limiting potentials plotted against the DFT-calculated adsorption free energy of OH$^*$ relative to Pt(111). \textbf{c,} Probabilistic activity volcano using $\Delta G_{OH^*}$, which we identify as the descriptor that maximizes the prediction efficiency. \textbf{d,} Prediction efficiency as a function of the activity interval of interest, $|\mathrm{U_{L,max}-U_{T}}|$. It can be seen that $\Delta G_{OH^*}$ is the descriptor that maximizes predictability for 4e$^-$ ORR.}
	\label{fig:ORR_4e}
\end{figure}

We follow the associative mechanism for the 4e$^-$ reduction\cite{norskov2004origin} and the electrocatalytic activity for ORR is determined by the free energies of adsorbed OOH*, OH* and O*.  However, the presence of scaling\cite{abild2007scaling} between these intermediates allows us to use a single descriptor for the activity\cite{calle2012first,man2011universality, stephens2012understanding}. We apply the probabilistic approach to predict ORR activity and consider various metallic facets. Scaling relations between the adsorbates (fig. \ref{fig:ORR_4e}(a) and fig. S6) allow us to describe the limiting potential as a function of one descriptor (fig. \ref{fig:ORR_4e}(b)). Let us begin with the case of the free energy of OH*, $G_{OH^*}$ being the descriptor, where the limiting potential is given by  $U_L = f(\Delta G_{OH^*}) = min(G_{OH^*}, 4.92-(3.11+\Delta G_{OH^*}))$. Hence the limiting potential is expressed as $U_L = min(G_{OH^*}, 4.92-(3.11+\Delta G_{OH^*}))$ V. The descriptor uncertainty is approximated based on the combined distribution of the surfaces explored \cite{deshpande2016quantifying}. The uncertainty in the scaling relation is incorporated by considering an ensemble of activity volcano relationships mapped from the ensemble of scaling relation intercepts. For each member of the ensemble of volcano relationships, the descriptor uncertainty is propagated. This allows us to compute the expected activity for each ensemble member and a probability-weighted average gives the activity PDF and the expected limiting potential (fig. \ref{fig:ORR_4e}(c)). In a similar manner, we construct the PDF maps of ORR activity and the corresponding expected limiting potentials using $\Delta G_{OH^*}$ and $\Delta G_{O^*}$ as the descriptors (fig. \ref{fig:ORR_4e}(d) and \ref{fig:ORR_4e}(e)). For the three descriptors, we can compare the predictability of activity based on the prediction efficiency. We find that the prediction efficiency follows the trend, $\eta_{pred}^{OH^*} > \eta_{pred}^{OOH^*} >  \eta_{pred}^{O^*} $, identifying that $\Delta G_{OH^*}$ is the descriptor for maximal predictability. It is worth highlighting that the use of $\Delta G_{O^*}$ as the descriptor uses two scaling relations while $\Delta G_{OH^*}$ and $\Delta G_{OOH^*}$ use only one, leading to improved prediction efficiency.

\begin{figure}[H]
	\centering
	\includegraphics[width=\textwidth]{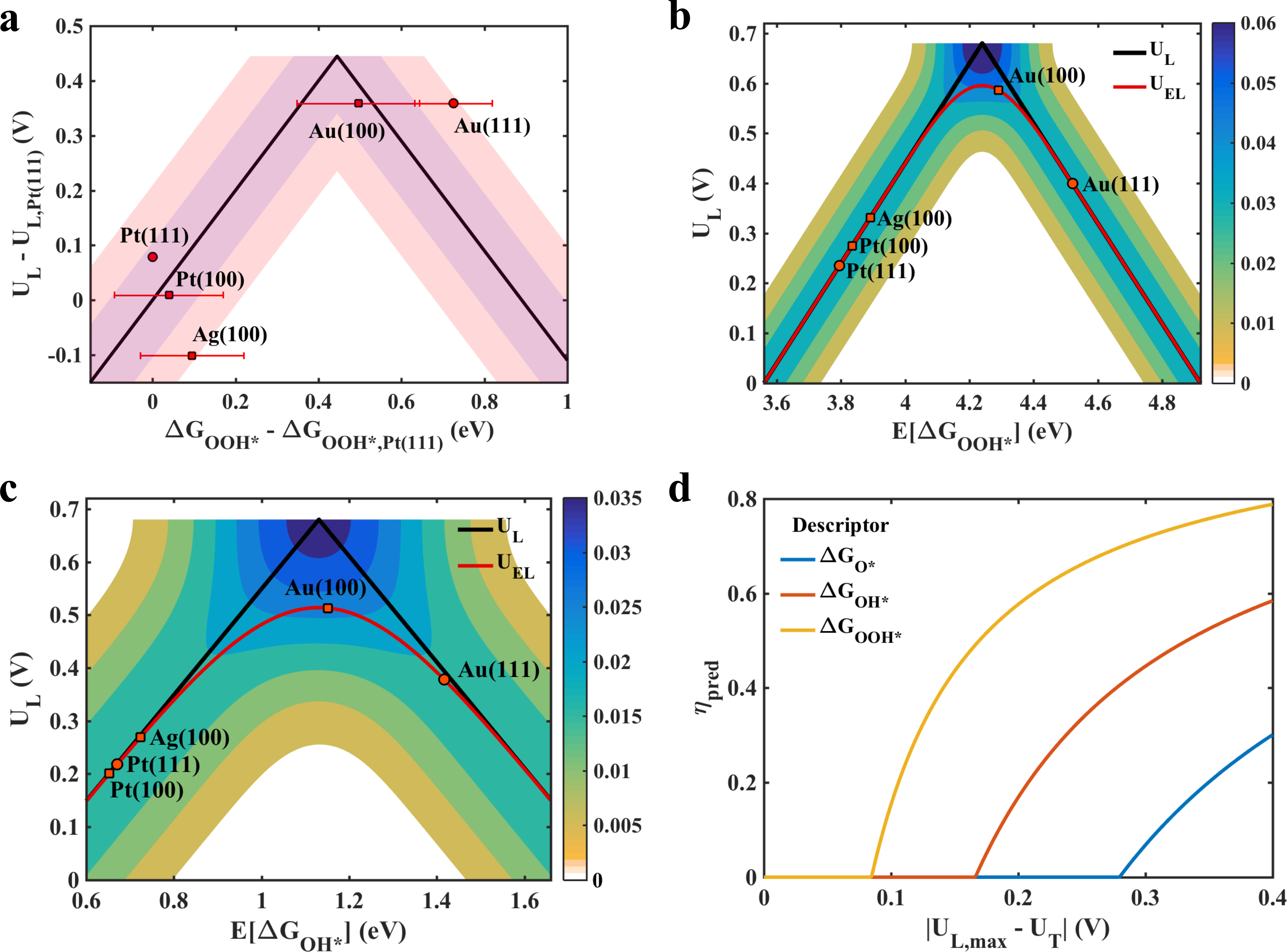}
	\caption{\textbf{Activity prediction for oxygen reduction to hydrogen peroxide using the probabilistic formalism. a,} Predicted activity (black lines) for 2e$^-$ ORR incorporating uncertainty plotted against the DFT-calculated adsorption free energy of OOH$^*$. The shaded regions depict one and two standard deviations for the predicted activity, and the red dots represent experimentally measured activity. Probabilistic activity volcano with \textbf{b,} $\Delta G_{OOH^*}$ and \textbf{c,} $\Delta G_{OH^*}$ as the descriptor. The activity volcano can also be computed using $\Delta G_{O^*}$ as the descriptor (fig. S9). \textbf{d,} Comparing the prediction efficiencies of the three descriptors as a function of the activity interval of interest, $|\mathrm{U_{L,max}-U_{T}}|$. This confirms that $\Delta G_{OOH^*}$ maximizes prediction efficiency for material screening.}
	\label{fig:ORR_2e}
\end{figure}

Oxygen can be electrochemically reduced through a 2e$^-$ process to produce hydrogen peroxide\cite{drogui2001hydrogen}. The reaction mechanism involves two concerted proton-electron transfers with a single intermediate OOH$^*$ \cite{viswanathan2012unifying}, which implies that an obvious choice of descriptor for activity prediction is $\Delta G_{OOH^*}$. However, using the scaling relation, $\Delta G_{OH^*}$ could alternatively be employed as a descriptor\cite{viswanathan2012universality}. It is worth highlighting that $\Delta G_{OOH^*}$ has been used as a descriptor to identify Hg-based alloys\cite{siahrostami2013enabling,verdaguer2014trends}.  The activity for H$_2$O$_2$ formation is a function of the binding energy of OOH$^*$, given by $U_L = f(\Delta G_{OOH^*}) = min(\Delta G_{OOH^*} - \Delta G_{H_2O_2} \ , \ \Delta G_{O_2} - \Delta G_{OOH^*})$. We first consider $\Delta G_{OOH^*}$ as the descriptor and propagate the uncertainty to the activity (fig. \ref{fig:ORR_2e} (b)). Using free energies of other intermediates as descriptors involves propagating the uncertainties in the descriptor and the scaling relation. Using this approach, the PDF map of the limiting potential using $\Delta G_{OOH^*}$, $\Delta G_{OH^*}$ and $\Delta G_{O^*}$ as the descriptors are shown in fig. \ref{fig:ORR_2e}(b), \ref{fig:ORR_2e}(c) and S9 respectively. We quantitatively show that $\Delta G_{OOH^*}$ maximizes predictability (fig. \ref{fig:ORR_2e}(d)) for identifying active materials for direct electrochemical H$_2$O$_2$ production.

\subsection{Oxygen Evolution Reaction}

\begin{figure}[H]
	\centering
	\includegraphics[width=\textwidth]{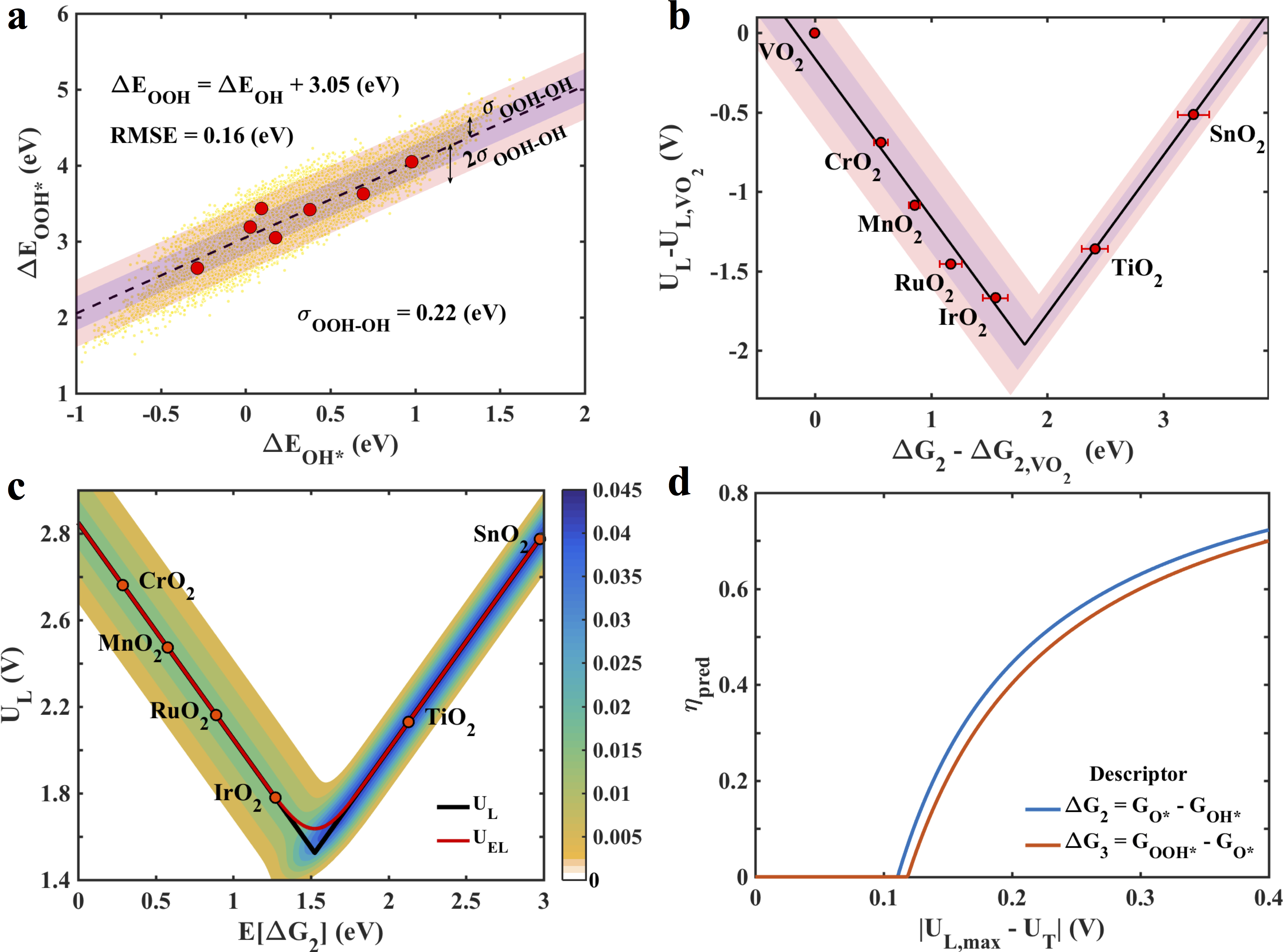}
	\caption{\textbf{Activity for oxygen evolution predicted using the identified descriptor by employing the probabilistic uncertainty propagation framework. a,} The computed scaling relation between the binding energies of the intermediates, OH* and OOH*, with the associated uncertainty in the intercept. \textbf{b,} The limiting potential as a function of the descriptor. $\Delta G_2 = G_{O^*} - G_{OH^*}$ . The shaded regions represent the uncertainty and the error-bars convey the descriptor uncertainty referenced to VO$_2$. \textbf{c,} Probability map of the limiting potential as a function of the expectation value of the descriptor ($E[\Delta G_{2}]$). The red curve represents the predicted expected activity, $U_{EL}$ \textbf{d,} Comparison of the prediction efficiencies of the two descriptors. Among the two, $\Delta G_2$ provides higher distinguishability of materials, by enhancing the prediction efficiency for oxygen evolution.}
	\label{fig:OER}
\end{figure}

Oxygen evolution is carried out under harsh oxidizing conditions and is a crucial process for the solar fuel generation.\cite{lewis2006powering} The foundational understanding of the OER on metal oxides has been built largely through experimental measurements and by empirically correlating them to the enthalpy of lower to higher oxide transition.\cite{trasatti1984electrocatalysis, trasatti1980electrocatalysis}  The insights developed using this approach have been limited due to the inability to accurately measure chemisorption energies.  This is due to the difficulty associated with preparing well-ordered single-crystalline oxides and the limited electrical conductivity, making them challenging for surface characterization.\cite{campbell2013enthalpies}

On the theoretical front, progress in understanding electrocatalysis has been limited by the accuracy of DFT in describing correlation in transition metal oxides.\cite{cohen2008insights,wang2006oxidation}  Using appropriate reference schemes, it has been shown that the formation energies of rutile oxides can be described well using DFT at the GGA level.\cite{martinez2009formation}  Man et al. explored trends in reactivity for oxygen evolution on rutile and perovskite surfaces, showing the existence of scaling relations.\cite{man2011universality}  Based on this analysis, they argued that $\Delta G = \Delta G_{O^*} - \Delta G_{OH^*}$ can be used as a descriptor for predicting activity.   Subsequently, two independent descriptors were used to predict the overpotential.\cite{seitz2016highly}  Despite these advances, the selection of the right descriptor for oxygen evolution remains elusive.   We fill this gap by exploring the associative mechanism for oxygen evolution on rutile oxide (110) surfaces.  The OER activity is determined by the adsorption energy of the reaction intermediates. We find that the scaling between the adsorption energies of OOH$^*$ and OH$^*$ has a slope close to 1 and the intercept is found to be 3.05 (Fig. \ref{fig:OER}(a)) \cite{man2011universality,viswanathan2014unifying}. The variation in the limiting-potential is therefore determined by $ \Delta G_{O^*}$. Hence, we can use $\Delta G_2 = (\Delta G_{O^*} - \Delta G_{OH^*})$ or $\Delta G_3 = (\Delta G_{OOH^*} - \Delta G_{O^*})$ as a descriptor. This implies that $U_L = \max(\Delta G_2, \ 3.05 - \Delta G_2) = \max(\Delta G_3, \ 3.05 - \Delta G_3)$. Following a similar approach to ORR, we show a probabilistic activity plot as a function of descriptors, $E[\Delta G_2]$ and $E[\Delta G_3]$, in figures \ref{fig:OER}(c) and S11. We quantitatively demonstrate that $\Delta G_2$ is the optimal descriptor for OER based on the prediction efficiency (Fig. \ref{fig:OER}(d)).

\subsection{Comparison between electrochemical reactions and approaches to improve prediction efficiency}
\begin{figure}[H]
	\centering
	\includegraphics[scale=0.6]{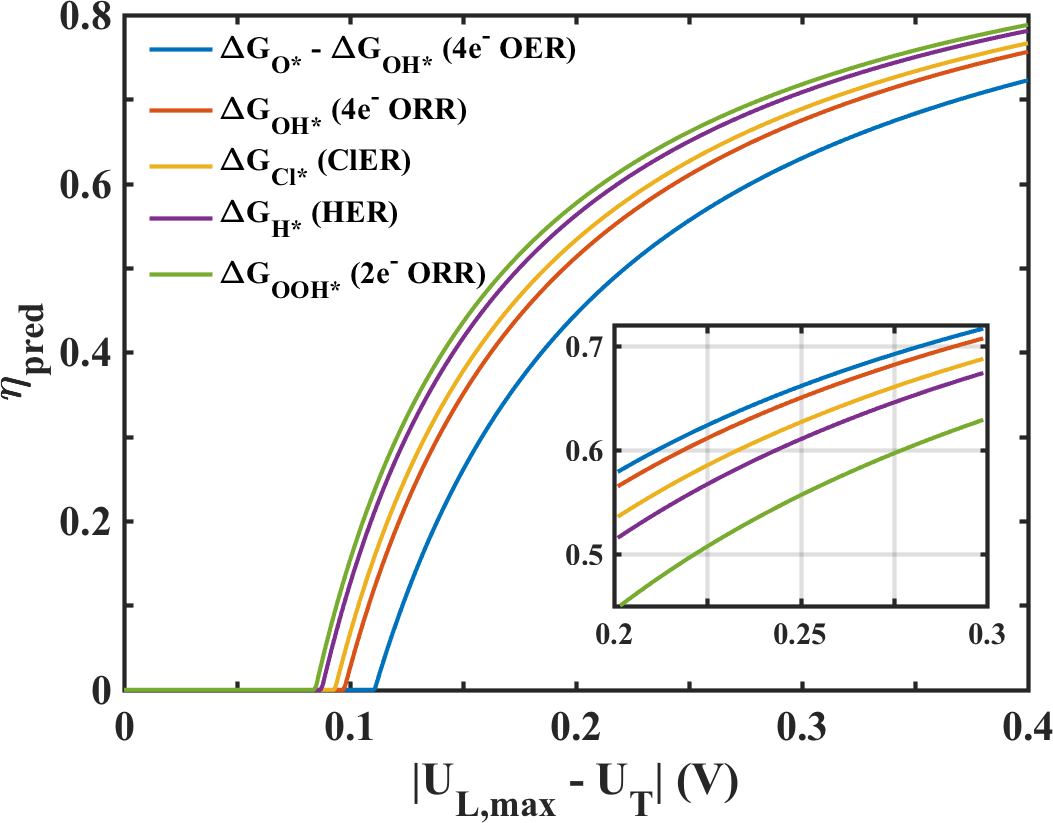}
	\caption{\textbf{Comparing the prediction efficiency of different electrochemical reactions.} Prediction efficiency, $\eta_{pred}$ as a function of $\mathrm{|U_{L,max}-U_{T}|}$, which represents the activity interval of interest for screening. It can be quantitatively seen that the ability of DFT to delineate materials decreases with increasing number of elementary electrochemical steps involved. Identifying highly active materials for oxygen reduction (4e$^-$) and oxygen evolution, therefore, requires a strong computation-experiment screening effort.}
	\label{fig:compare}
\end{figure}

The developed approach allows a quantitative comparison of the prediction efficiency across different electrochemical reactions.  We show a plot of the prediction efficiency for the optimal descriptor as a function of the overpotential for the considered electrochemical reactions in Fig \ref{fig:compare}.   Based on this analysis, we find that the prediction efficiency for 2e$^-$ electrochemical reactions such as HER and ClER is greater than that for 4e$^-$ ORR and OER.  This suggests that the likelihood of utilizing DFT calculations to identify highly-active candidates will be more probable for 2e$^-$ reactions compared to 4e$^-$ reactions.  Further, the differences in prediction efficiency between 2e$^-$ and 4e$^-$ suggest that predicting selectivity trends is fraught with challenges and requires a revisiting of the utilized descriptor that aims to optimize prediction efficiency for selectivity.  

We end with a word of caution regarding the selection of descriptors for determining trends in selectivity.  In this case, we need to identify a descriptor that can optimize the prediction efficiency for selectivity, not to optimize prediction efficiency for an individual electrochemical reaction.  The descriptor choice must be carried out through a mapping over maximal distiguishability in the selectivity space, which is a function of limiting potential for all the possible electrochemical reactions.  This involves taking the Lebesgue measure over the appropriate dimension, area in the case of selectivity between 2 reactions, volume in the case of selectivity between 3 reactions, etc.

In this work, although we have obtained the descriptor PDF using the BEEF-vdW XC functional, the probabilistic uncertainty-propagation framework is universally applicable regardless of how the descriptor PDF is obtained.  For example, the descriptor PDFs could be obtained from machine learning based models.\cite{Ling2017}

The descriptor uncertainty ($\sigma$), determines prediction efficiency, which could be reduced with higher order DFT methods. mBEEF is a meta-GGA based functional, with built-in error estimation capability.\cite{PhysRevB.91.235201} This functional could give rise to lower descriptor uncertainty leading to improved prediction efficiency. We suggest two approaches without increased computational complexity, through the use of (i) hybrid material reference,  and (ii) hybrid descriptors. We demonstrate in the Supporting Information that a two-material reference scheme for oxygen reduction leads to increased prediction efficiency.  Specifically, we show that referencing relative to a combination of Pt and Au leads to improved prediction efficiency.  Reference states leading to systematic error reduction is widely established for bulk formation energies.~\cite{martinez2009formation,christensen2015reducing} Descriptors of activity involving a linear combination of free energies of reaction intermediates involved can lead to higher predictability, but typically at the expense of interpretability.

\section{Conclusions} \label{Conclusions}

We have presented a method to carry out robust material selection through a systematic approach of incorporating uncertainty in density functional theory calculated energies. We argue that for increased prediction accuracy, screening studies should be based on the expected activity from the probabilistic approach. An important insight about the expected activity is that, when the descriptor distribution is experimentally trained as is the case with the BEEF-vdW XC functionality, the expected activity is a more accurate prediction. This implies that the mean value of the activity from a large number of experiments, which when enabled by high-throughput experimentation techniques, will better agree with the expected activity. This implies that identifying material candidates above the prediction limit requires more accurate computations and/or experiment-theory coupling. We define a quantity termed as the prediction efficiency which can be used to identify the optimal descriptor and compare the predictability of DFT across different electrochemical reactions. The prediction limit, which is the highest expected activity, represents the activity above which no assertive prediction can be made using computational approaches alone; any material that exhibits a higher activity experimentally is computationally indistinguishable. We demonstrate this by applying it to four of the most crucial electrochemical processes, namely hydrogen evolution, chlorine evolution, oxygen reduction and oxygen evolution.   We argue that the descriptor must be chosen such that it maximizes the prediction efficiency over the activity range of interest.  We show conclusively that the optimal descriptors for 4 and 2e$^-$ ORR are $G_{OH^*}$ and $G_{OOH^*}$, respectively.  Similarly, for oxygen evolution reaction, the optimal descriptor is identified to be $\Delta G_2 = \Delta G_{O^*} - \Delta G_{OH^*}$.   Finally, across reactions, we find that the prediction efficiency for 2e$^-$ electrochemical reactions such as HER and ClER is greater than that for 4e$^-$ ORR and OER.  This has important implications for analyzing trends in selectivity for an electrochemical reaction scheme.  Finally, we believe that the use of prediction efficiency should be ubiquitous and should form an integral part of descriptor-based activity predictions.

\section{Methods}
Calculations were performed using the projector augmented-wave (PAW) method as implemented in the GPAW program package using the recently developed Bayesian Error Estimation Functional with van der Waals correlation (BEEF-vdW), which has built-in error estimation capability \cite{wellendorff2012density}. The exchange correlation uses an ensemble of exchange correlation functionals resulting in an ensemble of energies from which the uncertainty in the adsorption energies can be calculated. For the hydrogen evolution reaction, metal catalysts of 2 x 2 surface cell with 4 layers separated by 10 $\mathring{A}$ of vacuum and periodic in x-y direction were considered. The hydrogen intermediate was adsorbed on an fcc(111) site with a coverage of 1/4 monolayer. A 10 $\times$ 10 $\times$ 1 k-point grid was used for the calculations. Rutile oxide catalysts were used for both the oxygen evolution reaction and the chlorine evolution reaction. For rutile oxides, we consider a 2$\times$1 surface unit cell and a 4$\times$4$\times$1 k-point grid. The surface of the unit cell contains two bridge and two cus sites. Adsorbates bind strongly on the bridge sites than on the cus sites and therefore the bridge site is always occupied with oxygen and inactive. All the OER and ClER intermediates were therefore adsorbed on the cus site. We consider a 1/2 monolayer (with respect to only the active cus sites) of the intermediates on the surface for both the reactions. Metal catalysts are used for the oxygen reduction reaction. Intermediates $\mathrm{OH^*}$ and $\mathrm{OOH^*}$ are modeled by including an explicit layer of water to account for hydrogen bonding on a 4-layered $\sqrt{3}\times\sqrt{3}$ configuration for metals and $2\sqrt{3}\times2\sqrt{3}$ configuration for $\mathrm{Pt_3Ni(111)}$ with 1/3 monolayer (ML) coverage. $\mathrm{O^*}$ is modeled on a 4 layered $2\times2$ configuration for metals and $2\times3$ configuration for $\mathrm{Pt_3Ni(111)}$  in an fcc site with a 1/4 monolayer (ML) coverage. A $6\times6\times1$ k-point grid was used for the $2\times2\times4$ unit cell and the k-points are scaled according to the different unit cells used. For all the calculations the bottom two layers were kept fixed and the top two layers with the adsorbates were allowed to relax with a force criterion of < 0.05 eV / $\mathring{A}$. Dipole correction was implemented in all calculations with metal catalysts. Spin-polarized calculations were carried out wherever necessary.

\section{Acknowledgements}\label{sec:ack}
D.K. and V.V gratefully acknowledge funding support from the National Science Foundation under award CBET-1554273.  V. S. and V. V. acknowledge support from the Scott Institute for Energy Innovation at Carnegie Mellon University.  The authors acknowledge Dr. Isabela C. Man for sharing structure files related to oxygen evolution and chlorine evolution.

\bibliographystyle{achemso}
\bibliography{refs}



\section*{Supplementary material}

# Supporting Information for Maximal predictability approach for identifying the right descriptors for electrocatalytic reactions

Dilip Krishnamurthy, Vaidish Sumaria, Venkatasubramanian Viswanathan

September 21, 2017

## 1 Adsorption Energy Distribution

Using BEEF-vdW functional, ensemble of adsorption energies for various intermediates involved in generated. We use the following methodology for estimating the combined overall error in the adsorption energies for various intermediates as described in [1]. The methodology is explained using the adsorption energy of $H^*$ for hydrogen evolution as an example.

- First the ensemble of $H^*$ adsorption energies for a given metal "X" with respect to a reference system (one that minimizes the overall prediction error) is calculated. In the case of Hydrogen evolution, the reference chosen is Rh(111). This is given as:

$$E_H(X|Rh(111)) = E_H(X) - E_H(Rh(111))$$

- We then center the distribution around the mean, which is given as:

$$\overline{E_H(X|Rh(111))} = E_H(X|Rh(111)) - \langle E_H(X|Rh(111)) \rangle$$

- This is carried out for all the catalysts ("X") considered and the combined distribution is constructed. The standard deviation of this combined distribution is the overall error in the adsorption energies ($\sigma_H$)

## 2 Hydrogen Evolution Reaction

### 2.1 Reaction Mechanism

$H^+ + e^- + * \rightarrow H^*$

$H^+ + e^- + H^* \rightarrow H_2(g) + *$

### 2.2 Calculation Details

The calulations were done on a $2 \times 2$ surface cell with 4 layers separated by 10 Å of vacuum. the slab is periodic in x and y direction. The hydrogen is adsorbed on a fcc(111) site. We considered surface coverage of 1/4 and 1 monolayer. The bottom two layers were fixed and the top two layers with the adsorbates are allowed to relax. All the structures were converged with a force criterion <0.05 eV/Å. A $10 \times 10 \times 1$ k-point grid was used for the calculations. The adsorption free energy of hydrogen is given as:

$\Delta E_H = \frac{1}{n}(E(\text{surf} + nH) - E(\text{surf}) - \frac{n}{2}E(H_2(g)))$

$\Delta G_{H^*} = \Delta E_H + \Delta E_{ZPE} - T\Delta S_H = \Delta E_H + 0.24$ (eV)

where $n$ is the number of H atoms sued in the calculation; $n = 1$ represents a coverge of 1/4 and $n=4$ represents a coverage of 1. The limit where proton transfer is exothermic ($\Delta G_{H^*} < 0$), the rate constant is independent of $\Delta G_{H^*}$ and the surface coverage is high. The limit where proton transfer is endothermic ($\Delta G_{H^*} > 0$), the reaction is activated by at least $\Delta G_{H^*}$ and proton transfer becomes difficult because hydrogen becomes unstable on the surface. The exchange current $i_0$ can be expressed in terms of the free energy of adsorption of hydrogen and charge transfer coefficient $\alpha$ using the following expression [2]:

$$i_0 = -ek_0(1 + \exp(|\Delta G_{H^*}|/(\alpha kT)))^{-1}$$

The uncertainty in the descriptor $\Delta G_{H^*}$ is found using the standard deviation of the opf the combined ensemble distribution of the free energies found using the BEEF-vdW exchange correlation.

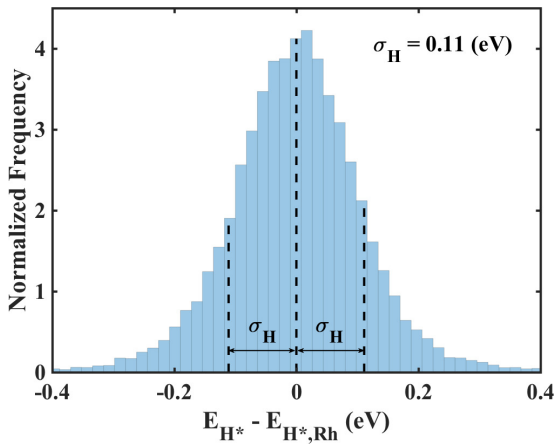

Figure S1: Plot of normalized frequency as a function of the adsorption energy of the intermediate H* relative to Rhodium. The standard deviation of this combined ensemble is $\sigma_H = 0.11$ eV.

Figure S1 shows the combined adsorption energy distribution for H* calculated on Au, Ag, Pd, Pt, Rh, Ir, Ni, W, Co, Cu, Mo, Re and Nb for fcc(111) facet. Its important to understand that the reference Rhodium is systematically chosen such that the the uncertainty in the descriptor ($\sigma_H$) is minimized.

## 2.3 Uncertainty Propagation

Following methodology is used to propagate the uncertainty in the descriptor to the exchange current:

- Using the uncertainty in the descriptor ($\sigma_H$) calculated using the combined re-centered distribution of adsorption energies, we now approximate a given computed $\Delta G_{H^*}$ as a normal distribution with its mean, $\mu = \Delta G_{H^*}$ and standard deviation of $\sigma_H$. This distribution can be expressed as:
$$X \sim \mathcal{N}(\mu, \sigma_H^2)$$

- Using this distribution the probability distribution of the descriptor, $\Delta G_{H^*}$, can be given by the Gaussian distribution:
$$p_x(x|\mu, \sigma_H^2) = \frac{1}{\sqrt{(2\pi\sigma_H^2)}} \exp\left(\frac{-(x-\mu)^2}{2\sigma_H^2}\right)$$

- We first sum over all the probability distributions of the descriptor that correspond to a particular exchange current value $i_0$. This is expressed as:
$$\hat{p}(i_0) = \int_{-\infty}^{+\infty} p_x(x)\delta(f(x) - i_0)dx$$

The dirac delta function ensures that we sum over all the descriptors value that correspond to a given exchange current value. This needs to be done for every value in the gaussian distribution $x$. The integral accounts for this complete distribution of the descriptor.

- The probability distribution of the exchange current can now be found by normalizing $\hat{p}$

$$p_{i_0}(i_0) = \frac{\hat{p}(i_0)}{\int_{-\infty}^{i_{0\max}} \hat{p}(i_0)}$$

- Expectation value, defined as a probability weighted average, is now obtained using the probability distribution of exchange current. The expectation value of the exchange current can be expressed as:

$$E(i_0) = \int_{-\infty}^{i_{0\max}} i_0 p_{i_0}(i_0) \, di_0$$

The probabilistic activity volcano and the expectation value for exchange current for transfer coefficient ($\alpha$) of 0.5 can be found in the main text and for transfer coefficient = 1 is shown in the figure S2.

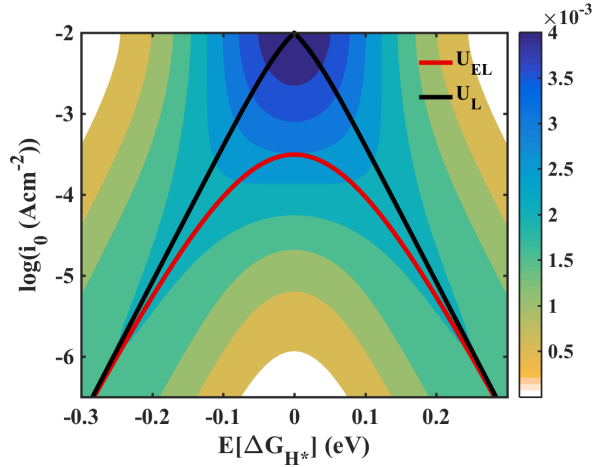

Figure S2: Probabalistic activity volcano for hydrogen evolution with transfer coefficient of 1. The solid black line represents the theoretical exchange current defined by the kinetic model and the red line represents the expectation value of exchange current.

In all the other reaction discussed further, the theoretical activity volcano is constructed on a thermodynamic analysis and not kinetic analysis. Hence in order to compare the prediction efficiency for Hydrogen evolution reaction with the other reactions, we construct a thermodynamic activity volcano and plot the limiting potential as a function of the adsorption free energy of hydrogen on the surface with its associated expectation value in figure S3

## 3 Chlorine Evolution Reaction

### 3.1 Reaction Mechanism

We use the Volmer-Heyvrosky reaction mechanism which ocuurs as follows:

$Cl^-(aq.) + * \rightarrow Cl^* + e^-$

$Cl^* + e^- + Cl^-(aq.) \rightarrow Cl_2(g) + 2e^- + *$

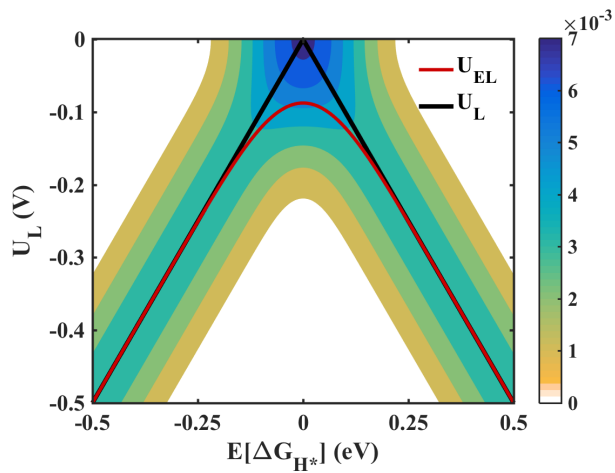

Figure S3: Thermodynamic activity volcano for Hydrogen evolution reaction showing the limiting potential for the reaction as a function of the adsorption free energy of hydrogen. The black line shows the theoretical limiting potential and the red line shows the expectation value of this limiting potential as a function of $\Delta G_{H^*}$

## 3.2 Calculation Details

The calculations were done on a periodically repeated 4 layered slab for the rutile (110) surfaces of $IrO_2$, $RuO_2$, $PtO_2$ and $TiO_2$. We consider a $2 \times 1$ surface unit cell and $4 \times 4 \times 1$ k-point grid. The bottom two layers were fixed and the top two layers with the adsorbates are allowed to relax. All the structures were converged with a force criterion $<0.05$ eV/Å. The surface of the unit cell contains two bridge and two cus sites. Adsorbates bind strongly on the bridge site than on the cus site and therefore the bridge site is occupied with oxygen. Hence we only focus on the cus sites. We use a 1/2 monolayer coverage (with respect to only the active cus site) of the intermediate on the surface. The adsorption free energy of chlorine is calculated as:

$$\Delta E_{Cl^*} = E(Cl^*) - E(*) - \tfrac{1}{2}E(Cl_2)$$

$$\Delta G_{Cl^*} = \Delta E_{Cl^*} + \Delta ZPE - T\Delta S = \Delta E_{Cl^*} + 0.37 \text{ (eV)}$$

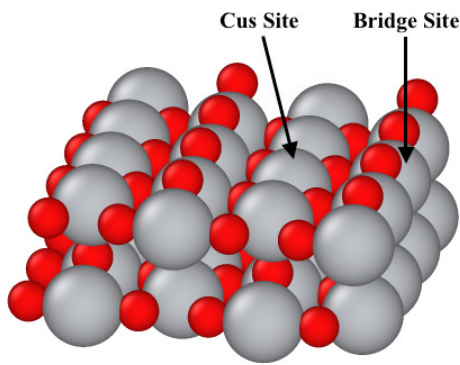

Figure S4: Visualization of the surface structure of rutile oxides (110) facet. Red and grey atoms represent oxygen and metal respectively. Bridges are inactive sites and are occupied with oxygen while the cus sites are the active sites

Figure S5 shows the combined adsorption energy distribution for H* calculated on $IrO_2$, $RuO_2$, $PtO_2$ and $TiO_2$ for (110) facet. $IrO_2$ is chosen as the reference because it minimizes the uncertainty in the descriptor ($\sigma_{Cl}$).

4

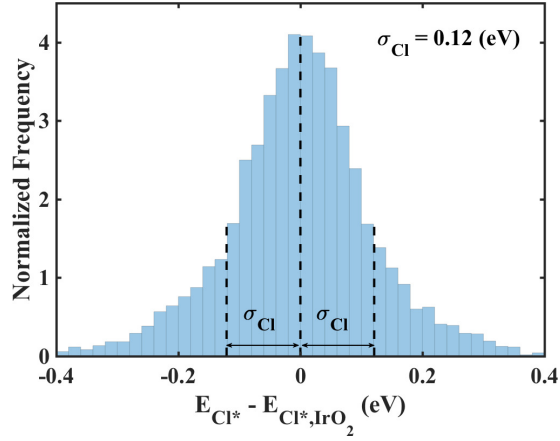

Figure S5: Plot of Normalized frequency as a function of the adsorption energy of the intermediate Cl* relative to $IrO_2$. The standard deviation of this combined ensemble is $\sigma_{Cl} = 0.12$ (eV).

## 3.3 Uncertainty Propagation

The methodology used for uncertainty propagation in this case is very similar to the one described for hydrogen evolution.

- The re-centered distribution of adsorption energies of Cl* is used to find the overall uncertainty in the descriptor ($\sigma_{Cl}$). Each of the calculated descriptor values ($\Delta G_{Cl^*}$) is assumed to be a normal distribution with the descriptor value as its mean $\mu = \Delta G_{Cl^*}$ and standard deviation of $\sigma_{Cl}$ which can be represented as:

$$X \sim \mathcal{N}(\mu, \sigma_{Cl}^2)$$

- Gaussian distribution is used to find the probability distribution of the descriptor $\Delta G_{Cl^*}$:

$$p_x(x|\mu, \sigma_{Cl}^2) = \frac{1}{\sqrt{(2\pi\sigma_{Cl}^2)}} \exp\left(\frac{-(x-\mu)^2}{2\sigma_{Cl}^2}\right)$$

- For a given value of limiting potential, we sum over all the probability distributions of the descriptors that map to that value of the liming potential. This is expressed as:

$$\hat{p}(U_L) = \int_{-\infty}^{+\infty} p_x \delta(f(x) - U_L) \, dx$$

- Normalized $\hat{p}$ defines the probability distribution of the liming potential as is expressed as:

$$p(U_L) = \frac{\hat{p}(U_L)}{\int_{-\infty}^{U_{L_{max}}} \hat{p}(U_L)}$$

- To calculate the expectation value of the limiting potential, a probability weighted average of limiting potential is calculated.

$$U_{EL} = \int_{-\infty}^{U_{L_{max}}} U_L \, p(U_L) \, dU_L$$

# 4 Oxygen Reduction Reaction

## 4.1 4e$^-$ Reaction Mechanism

We consider the following associative mechanism 4e$^-$ process for Oxygen Reduction reaction involving addition of a proton and an electron in each process.

$O_2 + 4H^+ + 4e^- + * \rightarrow OOH^* + 3H^+ + 3e^-$

$OOH^* + 3H^+ + 3e^- \rightarrow O^* + H_2O + 2H^+ + 2e^-$

$O^* + H_2O + 2H^+ + 2e^- \rightarrow OH^* + H_2O + H^+ + e^-$

$OH^* + H_2O + H^+ + e^- \rightarrow 2H_2O + *$

## 4.2 2e$^-$ Reaction Mechanism

Apart from the 4e$^-$ process, oxygen can also be reduced though a 2e$^-$ pathway in which the single intermediate, OOH$^*$, reduces to hydrogen peroxide H$_2$O$_2$. The associative 2$^-$ oxygen reduction proceeds as follows:

$O_2 + 2H^+ + 2e^- + {}^* \rightarrow OOH^* + H^+ + e^-$

$OOH^* + H^+ + e^- \rightarrow H_2O_2 + {}^*$

## 4.3 Details of the Calculations

Intermediates OH$^*$ and OOH$^*$ are modeled by including an explicit layer of water to account for hydrogen bonding on a 4 layered $\sqrt{3} \times \sqrt{3}$ configuration for metals and $2\sqrt{3} \times 2\sqrt{3}$ configuration for Pt$_3$Ni(111) with 1/3 monolayer (ML) coverage. O$^*$ is modeled on a 4 layered $2 \times 2$ configuration for metals and $2 \times 3$ configuration for Pt$_3$Ni(111) in a fcc site with a 1/4 monolayer (ML) coverage. A $6 \times 6 \times 1$ k-point grid was used for the $2 \times 2 \times 4$ unit cell and the k-points are scaled according to the different unit cells used. The bottom two layers were fixed and the top two layers with the adsorbates are allowed to relax with a force criterion of $< 0.05$ eV/Å. The adsorption energies od the various intermediates was calculated using the following equations:

$\Delta E_{O*} = E(O^*) - E(^*) - (E(H_2O) - E(H_2))$

$\Delta E_{OH*} = E(OH^*) - E(^*) - (E(H_2O) - 1/2\ E(H_2))$

$\Delta E_{OOH*} = E(OOH^*) - E(^*) - (2\ E(H_2O) - 3/2\ E(H_2))$

The entropy corrections and zero point energy corrections can be found in table S1. The gas phase values are from [3] and the values for the adsorbed species are taken from DFT calculations for O and OH adsorbed on Cu(111) from [4] and are assumed to be same for all the metals and the alloy. Gas-phase H$_2$O at 0.035 bar is used as the reference because at this pressure, gas-phase H$_2$O is in equilibrium with liquid water at 298 K. The entropy correction for the adsorbates on the surface are considered to be zero as the main contribution to the entropy is from the translational entropy.

Table S1: Zero point and entropic corrections at 298 K

|  | TS | T$\Delta$S | ZPE | $\Delta$ZPE | $\Delta ZPE - T\Delta S$ |
|---|---|---|---|---|---|
| H$_2$O(l) | 0.67 | 0 | 0.56 | 0 | 0 |
| OH$^* + 1/2$H$_2$ | 0.20 | -0.47 | 0.44 | -0.12 | 0.35 |
| O$^* +$ H$_2$ | 0.41 | -0.27 | 0.34 | -0.22 | 0.05 |
| $1/2$O$_2 +$ H$_2$ | 0.73 | 0.05 | 0.32 | -0.24 | -0.29 |
| H$_2$ | 0.41 | | 0.27 | | |
| $1/2$O$_2$ | 0.32 | | 0.05 | | |
| O$^*$ | 0 | | 0.07 | | |
| OH$^*$ | 0 | | 0.30 | | |

As discussed in section 1, the uncertainty in the adoption energy of the various intermediates is found from the combined adsorption energy distribution. The uncertainty in the adsorption energy of $O^*$ is $\sigma_O = 0.21$ (eV); uncertainty in the adsorption energy of $OH^*$ is $\sigma_{OH} = 0.09$ (eV) and uncertainty in the adsorption energy of $OOH^*$ is $\sigma_{OOH} = 0.11$ (eV).

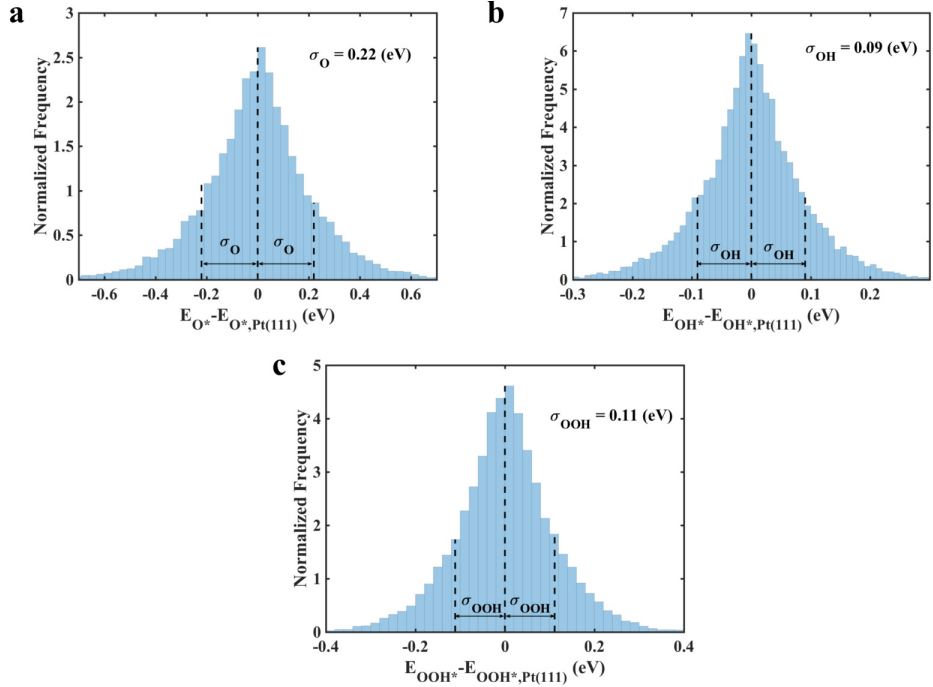

Figure S6: Normalized frequency as a function of the adsorption energy of the intermediate (a) $O^*$, (b) $OH^*$, (c) $OOH^*$ relative to Pt(111).

## 4.4 Uncertainty Propagation

### 4.4.1 4e$^-$ Reaction Mechanism

For all the metals facets that bond oxygen intermediate too strongly, the limiting potential $U_L$ can be given by:

$$U_L = \Delta G_{OH^*}$$

For the catalysts that bind oxygen intermediate weakly, the limiting potential is given by:

$$U_L = 4.92 - \Delta G_{OOH^*}$$

Exploiting the various scaling relations observed for the intermediates involved in the oxygen reduction reaction, we ca use $\Delta G_{O^*}, \Delta G_{OH^*}$ and $\Delta G_{OOH^*}$ as the descriptor to predict the limiting potential.

**(a) Choosing $\Delta G_{O^*}$ as the descriptor**

As the limiting potentials of the stronger binding and weaker binding legs of the volcano are given by $\Delta G_{OH^*}$ and $\Delta G_{OOH^*}$ respectively, we find a scaling relation between these two quantities as the chosen descriptor, $\Delta G_{O^*}$. The slope for this scaling is fixed to 0.5 and can be rationalized based on the bond conservation principles. The uncertainty in the scaling relation (of the form Y = 0.5*X + C) can be given as:

$$(\sigma_C)^2 = E[(Y - 0.5X)^2] - (E[Y - 0.5X])^2$$

$$(\sigma_C)^2 = E[(0.25X^2 + Y^2 - XY)] - (E[Y] - 0.5E[X])^2$$

7

$$(\sigma_C)^2 = 0.25E[X^2] + E[Y^2] - E[XY] - (0.25E[X])^2 - (E[Y])^2 + E[X]E[Y]$$
$$(\sigma_C)^2 = (0.25\sigma_X)^2 + (\sigma_Y)^2 - (E[XY] - E[X]E[Y])$$
$$(\sigma_C)^2 = (0.25\sigma_X)^2 + (\sigma_Y)^2 - (\mu_{XY} - \mu_X\mu_Y)$$

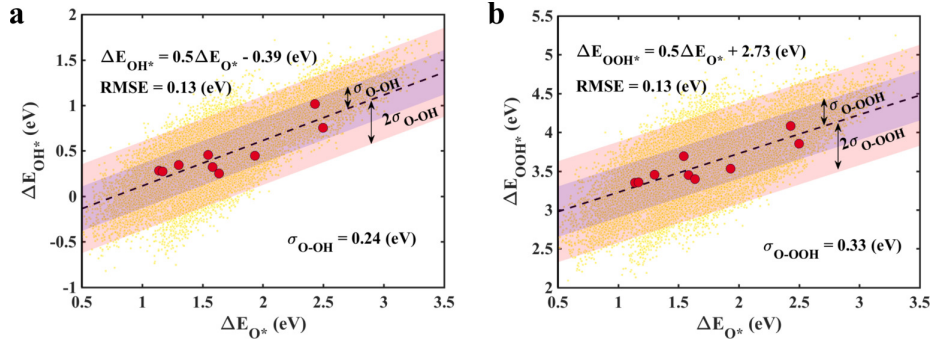

Figure S7: Scaling relation between (a) Adsorption Energy of $O^*$ and $OH^*$ and (b) Adsorption Energy of $O^*$ and $OH^*$

Using the scaling relation shown in figure S7, we can now define the limiting potential in terms of $\Delta G_{O^*}$ as follows:

- For stronger binding leg:

$$U_L = f(\Delta G_{O*}) = \Delta G_{OH^*} = 0.5\Delta G_{O^*} - 0.39$$

- For weaker binding leg:

$$U_L = f(\Delta G_{O*}) = 4.92 - \Delta G_{OOH^*} = 2.19 - 0.5\Delta G_{O^*}$$

The uncertainty in the stronger binding leg is defined by the uncertainty in the scaling between the adsorption energies of the intermediates $O^*$ and $OH^*$ ($\sigma_{O-OH}$) and the uncertainty in the weaker binding leg is defined by the uncertainty in the scaling relation between in the adsorption energies of the intermediates $O^*$ and $OOH^*$ ($\sigma_{O-OOH}$). Using this methodology, we use the following methodology to propagate the uncertainty of the descriptor and the scaling relation to the predicted limiting potential:

- Using the uncertainty in the descriptor ($\sigma_O$) calculated using the re-centered combined distribution of the adsorption energy of $O^*$, we assume each of the calculated descriptor $\Delta G_{O^*}$ values as normal distribution with its value as the mean ($\mu = \Delta G_{O^*}$) and the standard deviation as $\sigma_O$. This distribution is expressed as:

$$X \sim \mathcal{N}(\mu, \sigma_O^2)$$

- The probability distribution of the descriptor ($\Delta G_{O^*}$) normal distribution can be found using the gaussian distribution as:

$$p_x(x|\mu, \sigma_O^2) = \frac{1}{\sqrt{(2\pi\sigma_O^2)}} \exp\left(\frac{-(x-\mu)^2}{2\sigma_O^2}\right)$$

- $U_L$ function ($f(\Delta G_{O^*})$) should take into account the uncertainty in the scaling relation. Hence it can be better represented as:
  For stronger binding leg:

$$U_L = f(\Delta G_{O*}, K_s) = \Delta G_{OH^*} = 0.5\Delta G_{O^*} + K_s$$

For weaker binding leg:

$$U_L = f(\Delta G_{O*}, K_w) = 4.92 - \Delta G_{OOH^*} = K_w - 0.5\Delta G_{O^*}$$

8

where, $K_s = \mathcal{N}(\mu = -0.39, \sigma_{O-OH})$ and $K_w = \mathcal{N}(\mu = 2.19, \sigma_{O-OOH})$ are normal distributions. Hence for a given value of $\Delta G_{O^*}$, we have an ensemble of $U_L$ values which changes the picture of one activity volcano to an ensemble of activity volcanoes with different peaks. For computational purposes we assume this distribution to be discrete and define the random variables $k_s \in K_s$ and $k_w \in K_w$. The maximum limiting potential $U_{L_{max}}$ can be determined by solving the above mentioned two equations simultaneously. For a given activity volcano among the ensemble, at the descriptor value $\Delta G_{O^*} = (k_w - k_s)$, max limiting potential of $U_{L_{max}} = (k_s + k_w)/2$ is found. The uncertainty in the descriptor for each of these activity volcanoes is propagated to the limiting potential in a similar manner as described previously and then is averaged for all the activity volcanoes for a given descriptor value.

- Now for an $i^{th}$ activity volcano relationship in the ensemble we sum over all the probability distribution of the descriptor that map to that value of the limiting potential.

$$\hat{p}_i(U_L) = \int_{-\infty}^{+\infty} p_x \delta(f(x) - U_L) \, dx$$

- Normalized $(\hat{p})_i$ defines the probability distribution of the liming potential which is expressed as:

$$p_i(U_L) = \frac{\hat{p}_i(U_L)}{\int_{-\infty}^{(U_{L_{max}})_i} (\hat{p})_i(U_L)}$$

- To calculate the expectation value of the limiting potential, a probability weighted average of limiting potential is calculated.

$$U_{EL_i} = \int_{-\infty}^{(U_{L_{max}})_i} U_L \, p_i(U_L) \, dU_L$$

- This is done similarly for every member of the activity volcano ensemble and then is averaged over the ensemble for a given $\Delta G_{O^*}$.

Using this approach, we construct the probabilistic activity volcano with the corresponding expectation value of the limiting potential as shown in the figure S8.

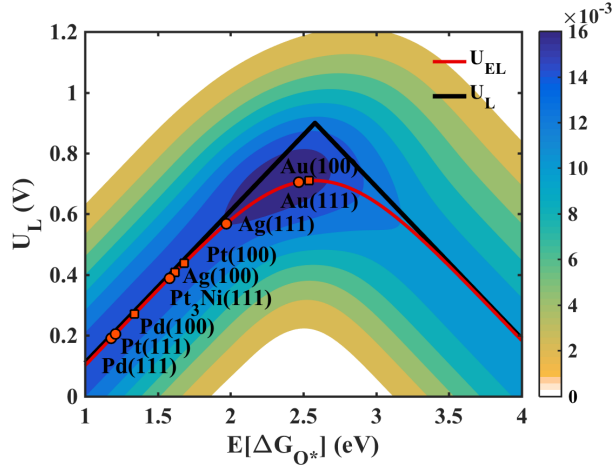

Figure S8: Probabilistic activity volcano for 4e$^-$ oxygen reduction reaction using $\Delta G_{O^*}$ as the descriptor. The Prediction Efficiency of this descriptor for the criterion of achieving limiting potential greater than that of Pt(111) is 0%.

**(b) Choosing $\Delta G_{OH^*}$ as the descriptor**

The uncertainty in the intercept of the scaling relation (which is of the form Y = X + C) between the adsorption energies of intermediate OH* and OOH* has been shown in the figure 3(a) can be found in the following way:

$$(\sigma_c)^2 = E[(Y - X)^2] - (E[Y - X])^2$$

$$(\sigma_c)^2 = E[(Y^2 + X^2 - 2XY)] - (E[Y] - E[X])^2$$

$$(\sigma_c)^2 = E[Y^2] + E[X^2] - 2E[XY] - (E[Y])^2 - (E[X])^2 + 2E[X]E[Y]$$

$$(\sigma_c)^2 = (\sigma_Y)^2 + (\sigma_X)^2 - 2(E[XY] - E[X]E[Y])$$

$$(\sigma_c)^2 = (\sigma_X)^2 + (\sigma_Y)^2 - 2(\mu_{XY} - \mu_X\mu_Y)$$

Using the scaling relation between the intermediates OH* and OOH* shown in the main text, we can define the limiting potential on terms of the chosen descriptor $\Delta G_{OH*}$ as follows:

- For the stronger binding leg:

$$U_L = f(\Delta G_{OH*}) = \Delta G_{OH*}$$

- For the weaker binding leg:

$$U_L = f(\Delta G_{OH*}) = 4.92 - \Delta G_{OOH*} = 1.81 - \Delta G_{OH*}$$

The uncertainty in the stronger binding leg is now defined by the uncertainty in the adsorption energy of the intermediate OH* ($\sigma_{OH}$) and the uncertainty in the weaker binding leg is defined by the uncertainty in the scaling relation between the intermediates OH* and OOH* ($\sigma_{OH-OOH}$). The following methodology was used to propagate the uncertainty in the descriptor and the uncertainty in the scaling relation to the limiting potential:

- Each of the calculated descriptor value $\Delta G_{OH*}$ is assumed as a normal distribution with its value as the mean and the standard deviation given by the uncertainty in the descriptor ($\sigma_{OH}$). This normal distribution can be represented as:

$$X \sim \mathcal{N}(\mu, \sigma_{OH})$$

- Using the gaussian distribution, the probability distribution of descriptor can be found

$$p_x(x|\mu, \sigma_{OH}^2) = \frac{1}{\sqrt{(2\pi\sigma_{OH}^2)}} \exp\left(\frac{-(x-\mu)^2}{2\sigma_{OH}^2}\right)$$

- In order to account for the uncertainty in the scaling relation, the function of limiting potential for the weaker binding leg (defined by scaling) can be better represented as:

$$U_L = f(\Delta G_{OH*}, K_w) = K_w - \Delta G_{OH*}$$

where $K_w = \mathcal{N}(\mu = 1.81, \sigma_{OH-OOH})$ is a normal distribution. Hence a given descriptor value generates an ensemble of predicted limiting potentials, giving rise to an ensemble of activity volcanoes. A discrete distribution of random variables $k_w \in K_w$ is generated to computationally simulate this problem. For a given activity volcano among the ensemble, the maximum limiting potential of $k_w/2$ is found for $\Delta G_{OH*} = k_w/2$. The probability distribution of limiting potential for a given activity volcano among the ensemble is found averaged over the whole ensemble for a given descriptor value as discussed for $\Delta G_{O*}$ as the choice of descriptor.

(c) Choosing $\Delta G_{OOH*}$ as the descriptor

The scaling relation relating the adsorption energy of OOH* and OH* can be exploited to define the limiting potential in terms of a single descriptor $\Delta G_{OOH*}$. Hence the limiting potential can be given as:

- For the stronger binding leg:

$$U_L = f(\Delta G_{OOH*}) = \Delta G_{OH*} = \Delta G_{OOH*} - 3.11$$

10

- For the weaker binding leg:

$$U_L = f(\Delta G_{OOH^*}) = 4.92 - \Delta G_{OOH^*}$$

Hence the uncertainty in the strong binding leg is defined by the uncertainty in the scaling relation and the uncertainty in the weaker binding leg is a function of the uncertainty of the descriptor. To propagate this uncertainty, we follow a similar approach as described for the $\Delta G_{OH^*}$ as the descriptor.

- Every calculated value of descriptor is assumed to be normal distribution with the $\mu = \Delta G_{OOH^*}$ and standard deviation of $\sigma_{OOH^*}$. This distribution is expressed as:

$$X = \mathcal{N}(\mu, \sigma_{OOH^*})$$

- The probability distribution of the descriptor can be expressed using the gaussian distribution as:

$$p_x(x|\mu, \sigma_{OOH}^2) = \frac{1}{\sqrt{(2\pi\sigma_{OOH}^2)}} \exp\left(\frac{-(x-\mu)^2}{2\sigma_{OOH}^2}\right)$$

- To incorporate the uncertainty in the scaling relation the limiting potential for the stronger binding leg can be represented as:

$$U_L = \Delta G_{OOH^*} - K_s$$

where $K_s = \mathcal{N}(3.11, \text{sigma}_{OH-OOH})$ is a normal distribution of the scaling intercept. Hence the uncertainty is propagated to an ensemble of activity volcanoes which is then averaged to find the probabilistic activity volcano with the expectation value of the limiting potential.

## 4.5 2e$^-$ Reaction Mechanism

The metals that bind oxygen intermediates too strongly, removal of OOH* is the potential determining step and the over potential is given by:

$$U_L = \Delta G_{OOH^*} - \Delta G_{H_2O_2}$$

The activity of the materials binding weakly to the catalysts is associated with the activation of $O_2$ and the limiting potential is given by:

$$U_L = \Delta G_{O_2} - \Delta G_{OOH^*}$$

Here the formation energies of hydrogen peroxide and oxygen is found using the thermodynamic tables as 3.56 eV and 4.92 eV respectively to avoid the well known issues related to the calculation of molecular reaction energies using DFT. The obvious descriptor for the activity would be $\Delta G_{OOH^*}$, but due to the known scaling between the adsorption energy of OH*-OOH* and O*-OOH*, $\Delta G_{OH^*}$ and $\Delta G_{O^*}$ can also be used the descriptors for the limiting potential.

**(a) Choosing $\Delta G_{O^*}$ as the descriptor**

The limiting potentials for both the weaker and stronger binding legs are expressed in terms of the chosen descriptor ($\Delta G_{O^*}$) using the scaling relation as:

- For stronger binding leg:

$$U_L = f(\Delta G_{O^*}) = \Delta G_{OOH^*} - 3.56 = 0.5\Delta G_{O^*} - 0.83$$

- For weaker binding leg:

$$U_L = f(\Delta G_{O^*}) = 4.92 - \Delta G_{OOH^*} = 2.19 - 0.5\Delta G_{O^*}$$

The uncertainty in the limiting potential is now associated with the uncertainty in the descriptor $\sigma_O$ as well as the uncertainty associated with the scaling relation $\sigma_{O-OOH}$. In order to propagate the error in both these quantities we follow the uncertainty propagation framework we discussed earlier

- Each of the calculated descriptor values $\Delta G_{O^*}$ is assumed to be a normal distribution centered around its value and having a standard deviation of $\sigma_O$. This distribution is now represented as:
$$X \sim \mathcal{N}(\mu, \sigma_O^2)$$

- The probability distribution of X can be obtained using the gaussian distribution as:
$$p_x(x|\mu, \sigma_O^2) = \frac{1}{\sqrt{(2\pi\sigma_O^2)}} \exp\left(\frac{-(x-\mu)^2}{2\sigma_O^2}\right)$$

- In order to account for the uncertainty in the scaling relation, as described earlier, we consider an ensemble of activity volcanoes and hence the limiting potential can be better represented as: For stronger binding leg:
$$U_L = f(\Delta G_{O^*}, K_s) = 0.5\Delta G_{O^*} + K_s$$

For weaker binding leg:
$$U_L = f(\Delta G_{O^*}, K_w) = K_w - 0.5\Delta G_{O^*}$$

where, $K_s = \mathcal{N}(-0.83, \sigma_{O-OOH})$ and $K_w = \mathcal{N}(2.19, \sigma_{O-OOH})$ are normal distribution defining the ensemble of activity volcanoes. For each member of the activity volcano, we propagate the uncertainty in the descriptor and find the probability distribution of limiting potential $p(U_L)$ and the expectation value of the limiting potential as

$$U_{EL} = \int_{-\infty}^{U_{L_{max}}} U_L \; p(U_L) \; dU_L$$

Unlike the case of 4e$^-$ process, where $U_{L_{max}}$ for each of the activity volcano in the ensemble is determined by the $K_{strong}$ and $K_{weak}$ distribution, for 2e$^-$ process, the maximum limiting potential is cut off at the equilibrium potential of 0.68 V for all the activity volcanoes. By taking an average over the entire ensemble, we find the overall expectation value of the limiting potential using $\Delta G_{O^*}$ as the descriptor and taking into account the uncertainty associated with both the descriptor and scaling.

The figure S9 shows the probabilistic activity volcano with the expectation value of the limiting potential for 2e$^-$ oxygen reduction reaction using $\Delta G_{O^*}$ as the descriptor.

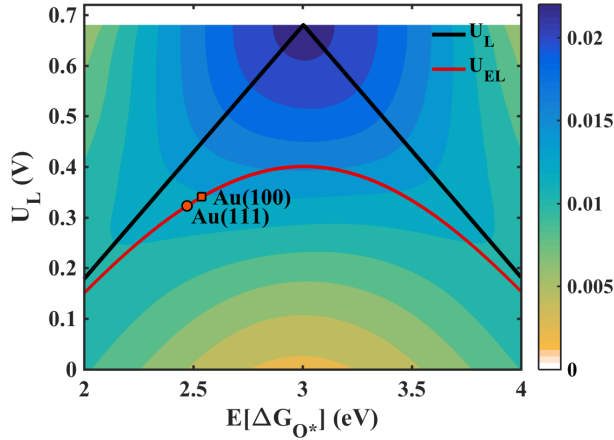

Figure S9: Probabilistic activity volcano for 2e$^-$ oxygen reduction reaction using $\Delta G_{O^*}$ as the descriptor. The large uncertainty in the descriptor value as well as the scaling relation results in reduced differentiability among materials based on the limiting potential.

(b) Choosing $\Delta G_{OH^*}$ as the descriptor

Using the scaling relation between the adsorption energy of the intermediates OH$^*$ and OOH$^*$, the limiting potential for 2e$^-$ ORR can be obtained by the single descriptor $\Delta G_{OH^*}$ as follows:

12

- For the stronger binding leg:

$$U_L = f(\Delta G_{OH^*}) = \Delta G_{OOH^*} - 3.56 = \Delta G_{OH^*} - 0.45$$

- For the weaker binding leg:

$$U_L = f(\Delta G_{OH^*}) = 4.92 - \Delta G_{OOH^*} = 1.81 - \Delta G_{O^*}$$

To propagate the uncertainty in the descriptor $\sigma_{OH}$ and the uncertainty in the scaling relation $\sigma_{OH-OOH}$, we use the same approach described for $\Delta G_{O^*}$ descriptor. The uncertainty in the descriptor is propagated by assuming the calculated descriptor value to be a normal distribution and the uncertainty in the scaling is propagated by considering an ensemble of activity volcanoes.

### (c) Choosing $\Delta G_{OOH^*}$ as the descriptor

Since an activity volcano for 2e$^-$ ORR with $\Delta G_{OOH^*}$ as the descriptor does not involve scaling, only the uncertainty in the descriptor needs to be propagated to the predicted limiting potential. The uncertainty propagation methodology will therefore be exactly the same as discussed for hydrogen and chlorine evolution reaction which involved a single intermediate in the reaction mechanism. Since there is not uncertainty with respect to scaling that needs to be considered, it is expected that $\Delta G_{OOH^*}$ behaves as the most efficient descriptor for 2e$^-$.

## 5 Oxygen Evolution Reaction

### 5.1 Reaction Mechanism

Oxygen Evolution reaction proceeds with the following associative mechanism:

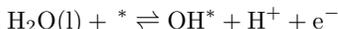
$H_2O(l) + {}^* \rightleftharpoons OH^* + H^+ + e^-$

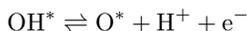
$OH^* \rightleftharpoons O^* + H^+ + e^-$

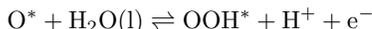
$O^* + H_2O(l) \rightleftharpoons OOH^* + H^+ + e^-$

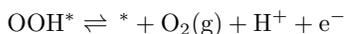
$OOH^* \rightleftharpoons {}^* + O_2(g) + H^+ + e^-$

The * represents an coordinately unsaturated site (cus) on the rutile oxide (110) surface as shown in S4.

### 5.2 Details of the Calculations

The calculations were done on a periodically repeated 4 layered slab for the rutile (110) surface of IrO$_2$, RuO$_2$, PtO$_2$, TiO$_2$, VO$_2$, CrO$_2$ and MnO$_2$. As discussed in section 3, we consider a $2 \times 1$ surface cell and $4 \times 4 \times 1$ k-point grid. We allow the system to relax keeping thr bottom 2 layers fixed and allow the top 2 layers with the adsorbates to move. The equations for the adsorption free energy of the intermediates remain same as that discussed in the section 4. Using the methodology discussed in section 1, we find the uncertainty in the adsorption free energy using the combined adsorption energy ensemble. The uncertainty in the adsorption energy of O$^*$ is $\sigma_O = $ (eV); uncertainty in the adsorption energy of OH$^*$ is $\sigma_{OH} = $ (eV) and uncertainty in the adsorption energy of OOH$^*$ is $\sigma_{OOH} = $ (eV).

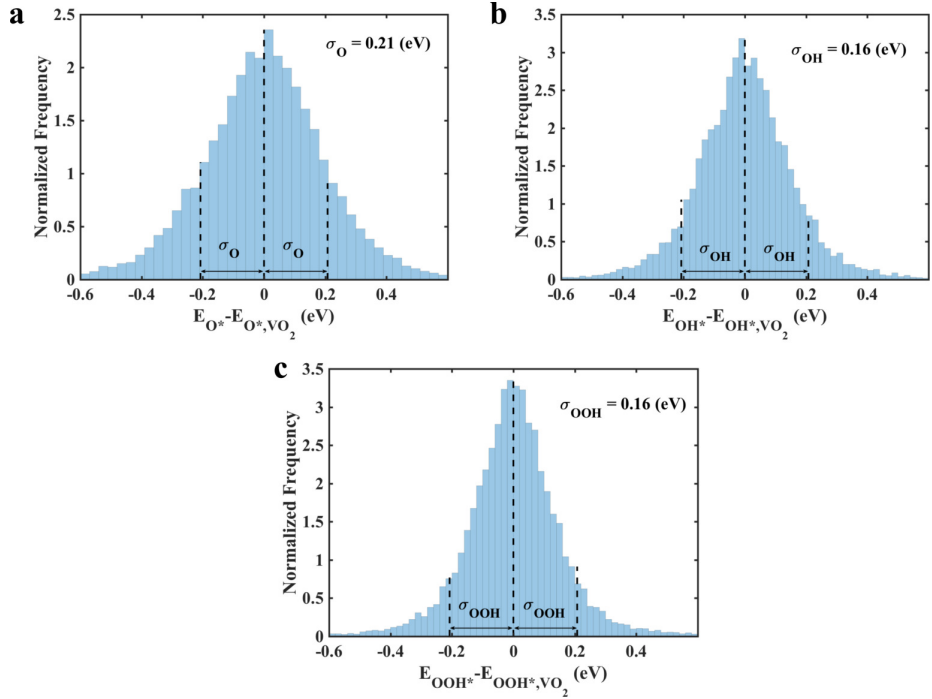

Figure S10: Normalized frequency as a function of the adsorption energy of the intermediate (a) O*, (b) OH*, (c) OOH* relative to VO$_2$.

The ZPE and entropy corrections can be found in table S2. For the adsorbed species the ZPE is obtained from [5], and was calculated for an adsorbate at the cut-site of RuO$_2$ and is considered to be same for each oxide.

Table S2: Zero point and entropic corrections at 298 K for rutile oxides (110)

|  | TS | T$\Delta$S | ZPE | $\Delta$ZPE | $\Delta ZPE - T\Delta S$ |
|---|---|---|---|---|---|
| H$_2$O(l) | 0.67 | 0 | 0.56 | 0 | 0 |
| OH* + 1/2H$_2$ | 0.20 | -0.47 | 0.50 | -0.06 | 0.41 |
| O* + H$_2$ | 0.41 | -0.27 | 0.34 | -0.22 | 0.05 |
| 1/2O$_2$ + H$_2$ | 0.73 | 0.05 | 0.32 | -0.24 | -0.29 |
| H$_2$ | 0.41 |  | 0.27 |  |  |
| 1/2O$_2$ | 0.32 |  | 0.05 |  |  |
| O* | 0 |  | 0.07 |  |  |
| OH* | 0 |  | 0.36 |  |  |

## 5.3 Uncertainty Propagation

For all the catalysts that bind the intermediates weakly, the limiting potential is given as:

$$U_L = \Delta G_2 = \Delta G_{O^*} - \Delta G_{OH^*}$$

For all the catalysts that bind the intermediates strongly, the limiting potential is expressed as:

$$U_L = \Delta G_3 = \Delta G_{OOH^*} - \Delta G_{O^*}$$

Using the scaling relation between the adsorption energies of the intermediates OH* and OOH* the limiting potential can be expressed in terms of a unique descriptor: $\Delta G_2$ or $\Delta G_3$. The descriptor that gives a higher prediction efficiency is the one desired to be used.

**(a) Choosing $\Delta G_2 = \Delta G_{O^*} - \Delta G_{OH^*}$ as the descriptor**

Using the scaling relation we can determine magnitude of the potential determining step ($G^{OER}$) as:

$$\begin{aligned} G^{OER} &= \max[\Delta G_2, \Delta G_3] \\ &= \max[(\Delta G_{O^*} - \Delta G_{OH^*}), (\Delta G_{OOH^*} - \Delta G_{O^*})] \\ &= \max[(\Delta G_{O^*} - \Delta G_{OH^*}), (3.05 - (\Delta G_{O^*} - \Delta G_{OH^*}))] \\ &= \max[\Delta G_2, 3.05 - \Delta G_2] \end{aligned}$$

Hence the limiting potential is given as:

- For stronger binding leg:
$$U_L = 3.05 - \Delta G_2$$

- For weaker binding leg:
$$U_L = \Delta G_2$$

Hence the uncertainty in the stronger binding leg is defined by the uncertainty in the scaling relation where as for the weaker binding leg is described by the uncertainty in the descriptor value. The uncertainty propagation method is similar to as described in oxygen reduction reaction with $\Delta G_{OH^*}$ as the descriptor. Following methodology was used:

- The calculated descriptor value $\Delta G_2$ is assumed to be normal distribution with the mean given by its value and the standard deviation of $\sigma_{G_2}$, where $\sigma_{G_2} = \sigma_O^2 + \sigma_{OH}^2 - 2(\mu_{O \times OH} - \mu_O \mu_{OH})$:

$$X \sim \mathcal{N}(\mu, \sigma_{OH})$$

- The probability distribution of the descriptor can be expressed using the gaussian distribution as:

$$p_x(x|\mu, \sigma_{G_2}^2) = \frac{1}{\sqrt{(2\pi \sigma_{G_2}^2)}} \exp\left(\frac{-(x-\mu)^2}{2\sigma_{G_2}^2}\right)$$

- The uncertainty in the scaling is incorporated by considering an ensemble of activity volcanoes as discussed earlier, hence the limiting potential of the stronger binding leg is now expressed as:

$$U_L = K_s - \Delta G_2$$

where $K_s = \mathcal{N}(3.05, \sigma_{OH-OOH})$ is a normal distribution of the scaling intercept. Uncertainty in each of the activity volcano in the ensemble is propagated in a similar way as described earlier and then averaged over the whole ensemble. The probabilistic activity volcano and the expectation value of the limiting potential obtained from this analysis is shown in the main text.

**(b) Choosing $\Delta G_3 = \Delta G_{OOH^*} - \Delta G_{O^*}$ as the descriptor**

Using the scaling relationship, we can determine the magnitude of the potential determining step ($G^{OER}$) in term of $\Delta G_3$ as follows:

$$\begin{aligned} G^{OER} &= \max[\Delta G_2, \Delta G_3] \\ &= \max[(\Delta G_{O^*} - \Delta G_{OH^*}), (\Delta G_{OOH^*} - \Delta G_{O^*})] \\ &= \max[(\Delta G_{O^*} - (\Delta G_{OOH^*} - 3.05)), (\Delta G_{OOH^*} - \Delta G_{O^*})] \\ &= \max[\Delta 3.05 - G_3, \Delta G_3] \end{aligned}$$

Using this, the limiting potential can be predicted using a single descriptor $\Delta G_3$ :

- For stronger binding leg:
$$U_L = \Delta G_3$$

- For weaker binding leg:
$$U_L = 3.05 - \Delta G_3$$

The uncertainty is propagated by assuming the descriptor value to be a normal distribution value with the mean given by it s value and the standard deviation of $\sigma_{G_3}$. This is represented as $X \sim \mathcal{N}(\mu, \sigma_{G_2})$. The uncertainty in the scaling is incorporated in a similar way as described earlier by considering an ensemble of activity volcanoes each corresponding to a different scaling intercept. Using this approach we construct the probabilistic activity activity volcano with $\Delta G_3$ as the activity descriptor and find the expectation value of the limiting potential as shown in S11

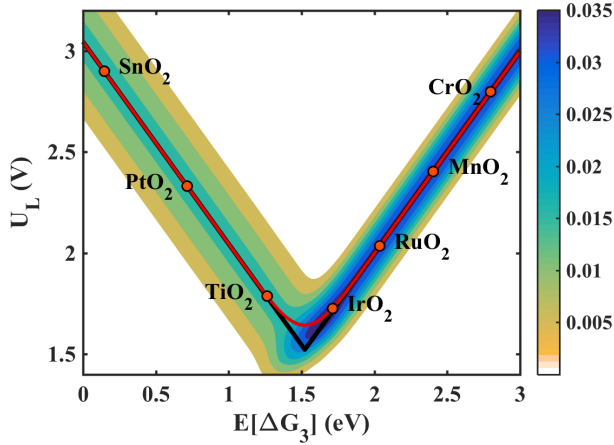

Figure S11: The probabilistic activity volcano for oxygen evolution reaction using $\Delta G_3 = \Delta G_{OOH*} - \Delta G_{O*}$ as the descriptor.

# 6 Approaches to improve Prediction Efficiency

As discussed in the paper, we see prediction efficiency can be improved by using (i) hybrid descriptors (ii) hybrid material references.

We showed earlier that $VO_2$ and Pt(111) reference minimized the overall uncertainty in the descriptor for oxygen evolution reaction and oxygen reduction reaction respectively. Here we show that using a combination of 2 descriptors for reference enables improving the prediction efficiency. Using a simple space search over the various combinations of the references, we find that a reference of $(0.4\Delta G_{2,TiO_2} + 0.6\Delta G_{2,RuO_2})$ for OER and $0.3\Delta G_{OH*,Pt(100)} + 0.7\Delta G_{OH*,Pd(111)}$ for ORR gives higher prediction efficiency.

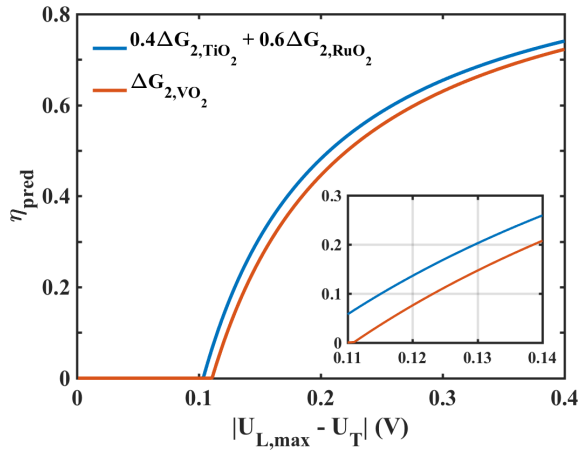

Figure S12: Comparing the prediction efficiency obtained using a single descriptor reference to a hybrid descriptor reference. From the plot it can observed that choosing a hybrid reference of $(0.4\Delta G_{2,TiO_2} + 0.6\Delta G_{2,RuO_2})$ instead of the single reference of $\Delta G_{2,VO_2}$ improved the prediction efficiency and the prediction limit.

16

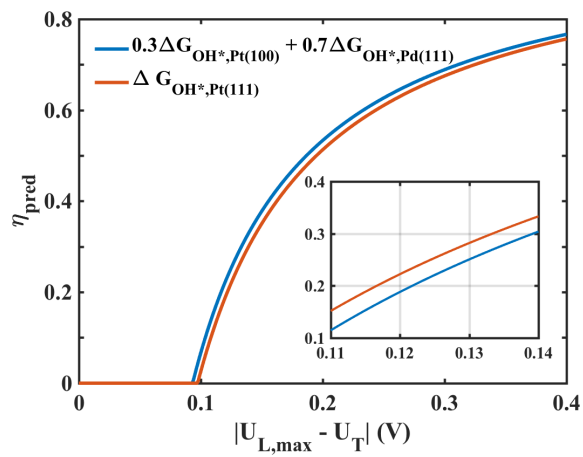

Figure S13: Comparing the prediction efficiency obtained using a single descriptor reference to a hybrid descriptor reference. From the plot it can observed that choosing a hybrid reference of $0.3\Delta G_{OH^*,Pt(100)} + 0.7\Delta G_{OH^*,Pd(111)}$ instead of the single reference of $\Delta G_{OH^*,Pt(111)}$ improved the prediction efficiency and the prediction limit.



\end{document}